\documentclass[prl,reprint,groupedaddress,english]{revtex4-1}
\usepackage{babel}
\usepackage{xcolor}
\usepackage{graphicx}
\usepackage{amsmath}

\definecolor{darkblue}{rgb}{0.1,0.2,0.6}
\definecolor{darkred}{rgb}{0.8,0.1,0.2}
\usepackage[colorlinks,citecolor=darkblue,linkcolor=darkred,urlcolor=darkblue]{hyperref}
\usepackage[all]{hypcap} % fix figure link problem. Links jump to top of figure now, not caption

\makeatletter
\adddialect\l@English\l@english
\makeatother

\newcommand{\bra}[1]{\langle\,#1\,|}
\newcommand{\ket}[1]{|\,#1\,\rangle}

\newcommand{\fullket}[1]{\left|#1\right>}
\newcommand{\fullbra}[1]{\left<#1\right|}

\begin{document}
\title{Universal behavior beyond multifractality in quantum many-body systems}
\author{David J. Luitz}
\affiliation{Laboratoire de Physique Th\'eorique, IRSAMC, Universit\'e de Toulouse, CNRS, 31062 Toulouse, France}
\author{Fabien Alet}
\affiliation{Laboratoire de Physique Th\'eorique, IRSAMC, Universit\'e de Toulouse, CNRS, 31062 Toulouse, France}
\author{Nicolas Laflorencie}
\affiliation{Laboratoire de Physique Th\'eorique, IRSAMC, Universit\'e de Toulouse, CNRS, 31062 Toulouse, France}
\email{nicolas.laflorencie@irsamc.ups-tlse.fr}
\date{October 9, 2013}

\pacs{75.40.Mg, 05.70.Jk, 75.10.Jm, 05.30.Rt, 89.70.Cf, 89.70.Eg}

\begin{abstract}
\makeatletter{}How many states of a configuration space contribute to a wave-function? Attempts to answer this
ubiquitous question have a long history in physics, and are keys to understand e.g. localization
phenomena. Beyond single particle physics, a quantitative study of the groundstate complexity for
interacting many-body quantum systems is notoriously difficult, mainly due to the exponential growth
of the configuration (Hilbert) space with the number of particles. Here we develop quantum Monte
Carlo schemes to overcome this issue, focusing on Shannon-R\'enyi entropies of groundstates of large quantum many-body systems. Our simulations reveal a generic multifractal behavior while the very nature of quantum phases of matter and associated transitions is captured by universal subleading terms in these entropies.
 
\end{abstract}
\maketitle
\makeatletter{}\makeatletter{}The introductory question arises naturally in different fields of research, notably in the study of Anderson localization~\cite{evers_anderson_2008,evers_fluctuations_2000}, complexity
theory~\cite{grassberger_complexity_1986}, quantum chaos~\cite{evangelou_2000}, diffusion-limited
aggregation~\cite{stanley_multifractal_1988}, quantum computing~\cite{georgeot_quantum_2000}
\textit{etc.}, whenever it is important to characterize how much of a configuration landscape
can be visited by a physical system. Measuring this complexity is crucial to establish the
efficiency of theoretical approximations ({\it e.g.} variational methods) and led to the
development of multifractal analysis~\cite{stanley_multifractal_1988, halsey_multifractal_1986}.
 
 To set up the problem, consider a normalized wave-function $\fullket{\Psi_0}$ of a quantum system 
expanded in a given orthonormal basis $\{\fullket{i}\}$ of size ${\cal N}$: $\fullket{\Psi_0} =
\sum_i \psi_i \fullket{i}$ with $\sum_i |\psi_i|^2 = 1$. The degree to which $\fullket{\Psi_0}$ is localized in $\{\fullket{i}\}$ can be quantified using the R\'enyi entropies \footnote{These entropies are closely related to the generalized inverse participation ratios $\sum_i p_i^{q}$, which are well-studied objects for instance in Anderson
localization~\cite{evers_anderson_2008,evers_fluctuations_2000}.}\begin{equation}
S_q = \frac{1}{1-q} \ln \sum_i p_i^q, \quad \mathrm{with} \quad q\in {\rm I\!R} \, \mathrm{ and } \,
p_i=|\psi_i|^2.
\label{eq:Sn}
\end{equation}
In the limit $q\rightarrow 1$, the Shannon entropy $S_1 =\lim_{q\rightarrow 1} S_q = - \sum_i p_i
\ln{p_i}$ is obtained. 

In the following, we denote $S_q$ and $S_1$ as the Shannon-R\'enyi (SR) entropies. 
Their scaling with system size provides a quantitative understanding of the wave-function localization~\cite{evers_anderson_2008,evers_fluctuations_2000} in configuration space: 
for \emph{single-particle} problems on a lattice (\textit{e.g.} Anderson localization), the size of the configuration
space $\cal N$ is proportional to the number of lattice sites.
In the thermodynamic limit ${\cal N} \rightarrow \infty$, the wave-function is
(i) localized if $S_{q}$ is bounded  by a constant,  (ii) delocalized if $S_q/\ln{\cal{N}}\to
1$,
(iii) multifractal if $S_q/\ln{\cal N}\to D_q$, where $D_q$ provide the fractal
dimensions
of the wave-function and a non-linear dependence on $q$ of $D_q$ marks the onset of
multifractality~\cite{janssen_multifractal_1994}, occurring for instance at the Anderson
transition~\cite{evers_anderson_2008}. Hence, the leading term in the scaling of SR entropies
contains the most relevant physical information in this case.

Characterizing the complexity of interacting {\it many-body} quantum systems with the same tools is
much more challenging,  as the number of configurations grows exponentially with the
number of particles ({\it e.g.} ${\cal N}=2^N$ for a collection of $N$ interacting two-level
systems).  In this work, we introduce two quantum Monte Carlo (QMC) schemes to calculate accurately
SR entropies for groundstates of large interacting quantum systems~\footnote{These methods are
efficient when the underlying QMC is, {\it i.e.} for models with no sign problem.}. 

Our results on one- ($d=1$) and two-dimensional ($d=2$) systems, as well as previous work on 
spin chains \cite{atas_multifractality_2012},  show that the notion of multifractality is of limited
use for many-body systems. Ground states of realistic many-body
Hamiltonians are found to generically exhibit $S_q = a_q N$ scaling at leading order, with $a_q$ a
non-trivial, non-linear function of $q$ that depends on microscopic details and is directly
proportional to the fractal dimension, \textit{e.g.} $a_q=D_q \ln 2$ for a spin $1/2$ wave-function
with maximal support. Therefore, \emph{multifractality} appears to be a \emph{generic feature} of many-body systems.

More strikingly, the results of our high-precision numerical simulations supported by
$d=1$ tractable examples
\cite{stephan_shannon_2009,stephan_renyi_2010,stephan_phase_2011,alcaraz_universal_2013} show that
\emph{subleading terms} $f_q(N)$ in the expansion 
$S_q(N)=a_q N+ f_q(N)$ contain \emph{universal}
information about many-body groundstates. In particular, we present several $d>1$ examples showing
the distinction between broken-symmetry phases and ``paramagnetic'' phases as well as the location of quantum phase transitions using sublinear terms $f_q(N)$ in the scaling of SR entropies.

{
We note at this stage the related progress made recently on the understanding of
condensed matter groundstates through their entanglement
properties~\cite{amico_entanglement_2008}. 
For instance, the study of
subleading terms in the scaling of entanglement entropies also allows a characterization of the
nature of these groundstates~(see e.g.
\cite{calabrese_entanglement_2004,
levin_detecting_2006,
kitaev_topological_2006,
isakov_topological_2011,
jiang_identifying_2012,
metlitski_entanglement_2011,
kallin_anomalies_2011,
kallin_entanglement_2013}). 
We emphasize however that SR entropies are {\it different} from entanglement entropies: 
they consider the system as a whole (not requiring a bipartition) and
characterize the complexity of the wave function in a given basis instead of its
entanglement.
Note that in some cases, the SR entropies of the groundstate of a system in dimension $d$ can be related to the entanglement entropies of a different wave function in {\it higher} dimension {\it d+1}~\cite{stephan_shannon_2009}.}

\begin{figure}
\centerline{\includegraphics[width=\columnwidth]{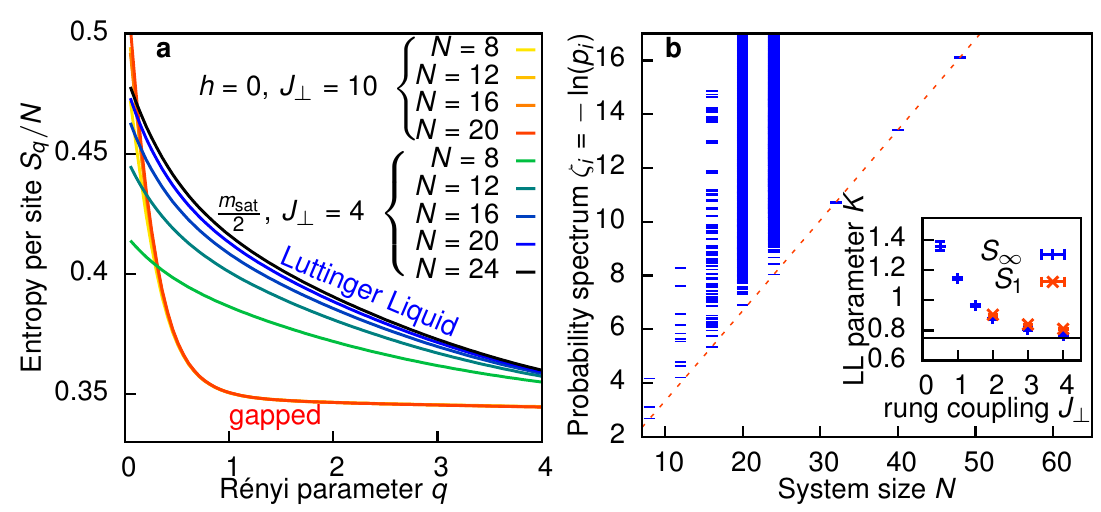}}
\caption{
a: SR entropies as a function of $q$ and system size $N$ for the groundstate of the Heisenberg
ladder at zero field (gapped) and at half saturated magnetization (LL) in the $S^z$ basis. As
$\lim_{N\rightarrow \infty}S_q/N=a_q$, the nontrivial behavior of $S_q/N$ signals multifractality.
b: Probability spectrum $\zeta_i$ at
$J_\perp=1.0$ and half saturated magnetization. For larger sizes only the most probable state was recorded. 
The dotted line is a
fit of $S_\infty$ to $a_\infty N + b_\infty + c_\infty/N$. 
Sizes $N=12$ and $N=20$ do not match the fit as the most probable state has periodicity 4 on each
leg and only clusters of size $N=8p$ with integer $p$ obey the condition. Incommensurate clusters
have a lower $p_\mathrm{max}$.
Error bars are reflected by line widths. The inset shows the $J_\bot$ dependence of the LL parameter
$K$ as obtained by fitting subleading terms in $S_1$ and $S_q$. The horizontal line displays
the limit $K\rightarrow 3/4$ for $J_\bot \gg 1$ \cite{giamarchi_coupled_1999}.}
\label{fig:ladder}
\end{figure}

\makeatletter{}
Studying subleading terms requires a large range of system
sizes $N$, a task rendered difficult by the exponential growth of configuration
space. A glance at Eq.~(\ref{eq:Sn}) suggests that we indeed need to compute all
$2^N$ coefficients $\psi_i$ in order to obtain $S_q$. Fortunately, this problem can be circumvented
by QMC techniques (see \textit{e.g.}~\cite{sandvik_computational_2010}) that
stochastically sum over configuration space using importance sampling. We now present 
two different QMC methods to compute $S_q$ for the GS wave function of generic many-body systems,
exploiting the fact that basis states $\fullket{i}$ are indeed sampled with the corresponding
probabilities $p_i$. This provides access to SR entropies of large quantum many body systems (up to
several hundreds of spins $S=1/2$, with Hilbert space dimensions larger than $10^{100}$) in
arbitrary dimension $d$ --- crucial for the study of subleading terms $f_q(N)$.

The \emph{ first method} measures $p_i$ directly. Path-integral QMC simulations or the related
Stochastic Series Expansion~\cite{sandvik_computational_2010} perform an importance sampling of the
partition function $Z={\rm Tr} \, \mathrm{e}^{-\beta H}=\sum_i \bra{i} \mathrm{e}^{-\beta H}
\ket{i}$
at finite temperature $T=1/\beta$ of a quantum system described by a Hamiltonian $H$. Observables $\langle {\cal O} \rangle = {\rm Tr}\, {\cal O} \hat{\rho}$ (with the density matrix $\hat \rho=e^{-\beta H}/Z$) can be easily obtained, in
particular when diagonal in the computational basis $\{\fullket{i}\}$. 
Indeed, one just needs to compute ${\cal O}$ in the configurations that appear in the $N_{\rm MC}$
states of the QMC sampling 
$ \langle {\cal O} \rangle \approx \frac{1}{N_{\rm MC}} \sum_{i=1}^{N_{\rm MC}} \langle i | {\cal O}
| i \rangle$.
We observe that the projector $\fullket{j} \fullbra{j}$ on the basis state
$\fullket{j}$ is diagonal in the computational basis, simply yielding $1$ if the state
$\fullket{i}$ found in the Markov chain of the QMC simulation is equal to $\fullket{j}$ and 
$0$ if $\fullket{i}\neq\fullket{j}$. 
Since $\hat{\rho}$ converges to the projector $|\Psi_0\rangle \langle \Psi_0 | $ into the GS of $H$ in the limit $\beta \rightarrow \infty$, the average probability to observe state
$\fullket{j}$ is given by $\langle p (\fullket{j}) \rangle \equiv \langle | j \rangle \langle j |
\rangle = \hat{\rho}_{jj} \stackrel{\beta \rightarrow \infty }{=} |\psi_j|^2,$
from which we can compute all $S_q$. Many interesting features can be obtained by sending
$q\rightarrow \infty$, for which we  simply record the probability $p_{\rm max}$ of observing the --
possibly degenerate -- most probable basis state: $S_\infty = -\mathrm{ln} \langle p_{\rm max} \rangle$.
The most frequent states are often found with a symmetry argument, or by a short QMC run. 

The \emph{ second method} computes the R\'enyi entropy $S_q$ for integer $q\geq 2$
using a replica trick: $q$ independent copies of $Z$ are simulated
simultaneously at low temperature to only sample the GS, thus performing the limit $\beta\rightarrow
\infty$
before studying the system size dependence. We perform measurements by checking whether
the QMC states $|j\rangle_q$ encountered for the $q$ copies are identical or not, defining
$P_\mathrm{identical}^q=1$ if all $|j\rangle_q$ states are the same, $0$ if not. As the copies are
independent, we have 
$\langle P_\mathrm{identical}^q \rangle = \langle \delta_{|j\rangle_1, | j
\rangle_2, \ldots | j \rangle_q} \rangle = \sum_j \hat{\rho}_{jj}^{q}  \stackrel{\beta \rightarrow \infty
}{=} \sum_j |\psi_j|^{2q},$
from which we can deduce $S_q$. As there is no need to save all
$p(|j\rangle)$, larger system sizes can be reached with this method. {This method can be
seen as a direct numerical implementation of the replica trick used in analytical calculations (see
\textit{e.g.} the ``book'' picture of Ref.~\cite{stephan_renyi_2010}) 
\footnote{The method is distinct from the Monte
Carlo method used in studies of entanglement entropies~\cite{hastings_measuring_2010,humeniuk_quantum_2012}: there is no ``swap'' of
configurations involved here, as all replicas are sampled independently according to the same density
matrix.}.
}

We provide a detailed discussion for the practical implementation including the exploitation of
symmetries in \cite{supplementary} 
and consider from now on the case of $N$ quantum spins $S=1/2$ and $q\geq 0$ only.

\begin{figure*}
\centerline{\includegraphics[width=14cm]{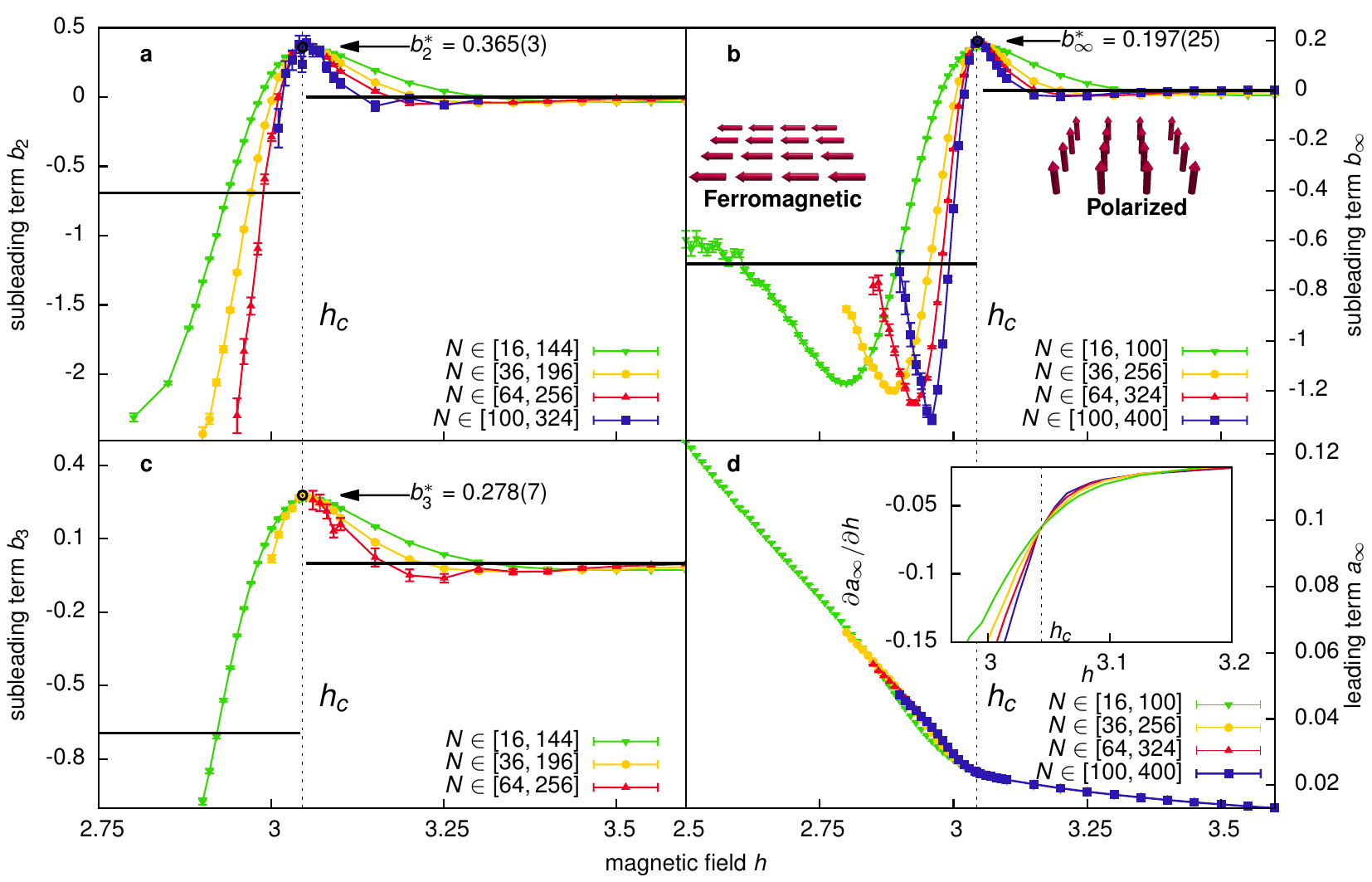}}
\caption{(a-c) Subleading coefficients $b_q$ of SR entropies in the $\sigma^z$ basis for the square
lattice Ising model in a transverse field $h$ for different $q$, as obtained from fits to $S_q=a_q N + b_q
+c_q/N + d_q/N^2$. We display fits over
different lattice size windows such as to assess the development towards the thermodynamic limit.
Subleading terms converge rapidly at criticality and display a marked jump as $h$ is tuned away from
its critical value. Straight lines show the limiting cases for high ($b_q=0$) and low field
($b_q=-\ln 2$). The convergence towards the low field thermodynamic limit $-\ln 2$ is not
visible here (except for $q=\infty$), due to limited accessible system sizes, as discussed in detail
in \cite{supplementary}.
(d) The field dependence of $a_\infty$ reveals a clear change of slope at $h_c$,
with a crossing point of $\mathrm{d}a_\infty/\mathrm{d}h$ for different finite-size fits (inset).}
\label{fig:ising}
\end{figure*}

Let us first apply these methods to $d=1$ quantum systems.
It is well established that low-energy physics of $d=1$ critical systems, such as spin chains or
quantum wires can be described by Luttinger Liquid (LL)
theory~\cite{giamarchi_quantum_2004}, with a key characteristic: the so-called LL parameter $K$.
Recently, St\'ephan {\it et al.}~\cite{stephan_shannon_2009,stephan_renyi_2010,stephan_phase_2011}
and Zaletel \textit{et al.}~\cite{zaletel_logarithmic_2011}, have highlighted the
connection between subleading terms in the scaling of SR entropies and $K$  for LL
systems. Using conformal field theory and numerics, Ref.~\cite{stephan_shannon_2009} has shown that
SR entropies of periodic spin chains of length $N$ admit the scaling behavior
\begin{equation}
S_q(N)=a_q N + b_q + {\cal{O}}\left(1/N\right)
\label{eq:simple_scaling}
\end{equation}
with $a_q$ the non-universal, model and $q$-dependent leading factor, whereas $b_q$
denotes the first subleading, constant with system size, coefficient in the expansion scaling.
$b_q$ is shown~\cite{stephan_shannon_2009,stephan_phase_2011} to be simply related to the
groundstate degeneracy for gapped systems, and for LL to $K$  by $b_q =-\frac{1}{2}(\ln K +\frac{\ln q}{q-1})$ (using the convention of Ref.~\cite{giamarchi_quantum_2004} for $K$).
This formula holds below a critical value $q_c=K {\cal D}^2$ 
for which a phase transition occurs~\cite{zaletel_logarithmic_2011,stephan_phase_2011} towards a phase
where $b_q$ is dominated by the most probable state with multiplicity ${\cal D}$:
$b_{q>q_c}=\frac{1}{1-q}(q\ln\sqrt{K}+ \ln{{\cal D}})$.

Non-trivial LL physics occurs for 2-leg spin ladder materials
\cite{dagotto_surprises_1996,giamarchi_coupled_1999}
in a magnetic field, as seen in recent nuclear magnetic resonance or inelastic neutron scattering
experiments~\cite{ruegg_thermodynamics_2008,klanjsek_controlling_2008}. Ladder systems are governed by the Hamiltonian 
\begin{equation*}
    {H}_{\rm lad} = \sum_{i,\alpha=1,2}\left( J_{\parallel}{\bf S}_{i,\alpha} \cdot {\bf S}_{i+1,\alpha}
    -h S_{i,\alpha}^z\right)
    + J_\bot  \sum_{i} {\bf S}_{i,1} \cdot {\bf S}_{i,2} 
\end{equation*}
with ${\bf S}_{i,\alpha}$ a spin-$1/2$ operator at site $i$ of leg $\alpha$, $J_{\parallel}=1$
($J_\perp>0$)  the antiferromagnetic couplings along legs (rungs), and $h$ the magnetic field. 
In the absence of an external field, the non-degenerate singlet
groundstate is separated from the first excited triplet state by a finite energy gap
$G\propto J_\perp$. Yet, when a sufficiently strong field $h>G$ is applied, an
effective LL description based on interacting hard-core bosonic triplet excitations
\cite{giamarchi_coupled_1999} is possible. As displayed in Fig.\ref{fig:ladder},
our QMC results demonstrate that the nature of the quantum physical state
is captured in the constant term $b_q$ of $S_q$, while $a_q$ vary non-trivially
with $q$ for both gapped and LL regimes. More precisely, at $h=0$ the non-degenerate gapped ground
state has a subleading term $b_q=0$ 
which can be analytically obtained in the large $J_\bot$ limit, leading to $S_q/N$ independent of
$N$ (Fig.\ref{fig:ladder}a). 
For $h>G$, in the LL regime, $b_q$ becomes a non-trivial function of $q,h,J_\perp$ from which the
parameter $K$ is extracted using the above expressions. 

\begin{figure}
\includegraphics[width=\columnwidth]{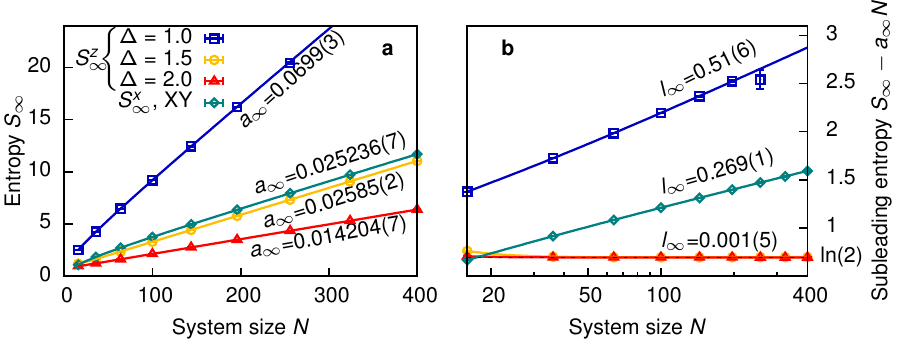}
\caption{a: $S_\infty$ for the $d=2$ isotropic ($\Delta=1$)
and anisotropic ($\Delta\neq 1$) XXZ model on the square lattice as a function of system size $N$. Lines are fits to the form
$S_\infty=a_\infty N + l_\infty \ln N + b_\infty + c_\infty/N + d_\infty/N^2$. b: Subleading logarithmic terms are highlighted in semi-log scale by subtracting $a_\infty N$ from $S_\infty$. Reported values of $a_\infty$ and $l_\infty$ are estimates from the best fit including error bars from a bootstrap analysis.}
\label{fig:XXZ}
\end{figure}

While we were able to record the entire spectrum of $p_i$ (represented in Fig.\ref{fig:ladder}b as
$\zeta_i=-\ln(p_i)$ \textit{vs.} $N$ for $J_\perp=1$) up to $N=24$ spins, the case $q\rightarrow \infty$
(requiring only one basis state),
permits dealing with larger systems: here up to $N=48$. 
Already with such relatively small lattices, one can precisely extract LL parameters for various
$J_\perp$ at finite external field $h$ such that the triplet density is half of its value at saturation.
QMC estimates for both $K=\exp(-(1+2b_1))$ and $K=\exp(-2b_\infty)$ are plotted \textit{vs.} $J_\perp$ as inset of Fig.\ref{fig:ladder}b, with a
better accuracy using $S_\infty$ due to larger accessible $N$. 
This result shows that the study of subleading terms in SR entropies together with our QMC method 
provides a very simple and efficient tool to compute the LL parameter (usually a
complicated task, requiring large $N$). Physically, it allows to observe a transition
from repulsive $K<1$ to the attractive regime $K>1$ around $J_\perp\simeq 1.5$, as discussed
recently for the ladder material $(\mathrm{C}_7 \mathrm{H}_{10} \mathrm{N} )_2
\mathrm{Cu} \mathrm{Br}_4$~\cite{hong_field-induced_2010,jeong_attractive_2013}.

In $d>1$ where the scaling of $S_q$ is essentially unknown due to size restriction, our QMC schemes access sufficiently large $N$ to allow us to address this question. We first
consider the $d=2$ transverse-field Ising model 
\begin{equation*}
H_{\rm Ising}= - \sum_{\langle i,j \rangle} \sigma_i^x \sigma_j^x - h \sum_i \sigma_i^z,
\end{equation*}
a paradigmatic example of a quantum phase transition
(QPT) between a low-field ferromagnetic phase (which breaks the $Z_2$ spin reversal symmetry) and a
high-field polarized phase. ${\mathbf \sigma_i^{x,z}}$ are Pauli matrices and $\langle i,j \rangle$
denote nearest-neighbor pairs on the square lattice.  The QPT occurs at $h_c \simeq 3.044$ on the
square lattice \cite{Blote}. For this {\it discrete symmetry breaking}, our QMC results for SR
entropies in the $\{\sigma^z\}$
 basis are very well-fitted by Eq.~(\ref{eq:simple_scaling}). 
 Fig.~\ref{fig:ising} reveals the {\it universal} nature of the subleading constant term $b_q$:
 throughout {\it all} the ferromagnetic phase, $b_q\rightarrow -\ln(2)$ in the thermodynamic limit
 (\textit{cf.} \cite{supplementary} for a discussion on finite-size effects), while $b_q \rightarrow 0$ in the polarized (paramagnetic) phase, and this {\it for all} $q>0$ considered.  These two constants are easily understood from the limiting cases $h=0$ and $h\rightarrow \infty$. Quite strikingly, $b_q (h_c) \rightarrow b_q^*$ at criticality, where $b_q^*$ is a non-trivial,
 $q-$dependent, constant.  We believe $b_q^*$ to be a {\it universal} function of $q$,
 characteristic of the $d=3$ Ising universality class, as corroborated by results for the same
 model on the triangular lattice (\textit{cf.} \cite{supplementary}). 
 Our data are compatible with $b_{q>1}^* = \frac{q}{q-1} b_\infty^*$ with
 $b_\infty^* \simeq 0.19(2)$, which indicates that the system is effectively locked in and physics dominated by the non-degenerate configuration with maximal probability $p_{\rm max}$ in the long wavelength limit~\cite{stephan_shannon_2009,stephan_phase_2011}. We expect the precise value of the leading term coefficient $a_q$ to be model-dependent, but observe nevertheless (Fig.~\ref{fig:ising}d) a clear signal of the QPT in its field dependence.

We finally highlight the different nature of subleading corrections in models that exhibit {\it
continuous symmetry} breaking in the thermodynamic limit. Consider the antiferromagnetic spin-$1/2$
XXZ model 
    $H_{\rm XXZ}= \sum_{\langle i,j \rangle} S_i^x S_j^x +S_i^y S_j^y + \Delta S_i^z S_j^z$
on the square lattice with the spin anisotropy $\Delta \geq 0$. For $0 \leq \Delta < 1$ (including
the XY model at $\Delta=0$), the groundstate breaks the continuous $U(1)$ rotation symmetry around
$z$ axis, while for $\Delta > 1$ a discrete Ising symmetry is broken. The isotropic point $\Delta=1$
(Heisenberg model) has an enhanced continuous $SU(2)$ spin rotation symmetry which is also broken in
the groundstate.
All broken continuous symmetries are associated with the presence of gapless Goldstone modes, while
for $\Delta>1$ the doubly-degenerate groundstate is separated by a gap from the first excitation. In the case
of continuous symmetry breakings, our QMC results (see Fig.~\ref{fig:XXZ} and \cite{supplementary})
demonstrate the presence of a {\it logarithmic subleading correction} in the scaling of SR entropies 
\begin{equation}
S_q(N)=a_q N + l_q \ln{N}+ b_q + \frac{c_q}{N} + \ldots, 
\label{eq:log_scaling}
\end{equation}
which are absent in the discrete symmetry breaking case, where we recover a 
constant term $b_q= \ln{2}$. We systematically found a logarithmic subleading term for all systems
with a continuous symmetry breaking that we have studied (\textit{cf.} \cite{supplementary} for results on models with a next-nearest neighbor coupling), at least for $q>1$. Similar
logarithmic corrections have been found in the scaling of entanglement entropies of systems with
continuous symmetry breaking, with a $q$-independent universal coefficient solely related to the
number of Goldstone modes~\cite{metlitski_entanglement_2011}. We observe in our simulations that
$l_q$ varies with $q$ (a scaling $l_{q>1}=\frac{q}{q-1}l_\infty$ seems to
hold), in contrast to entanglement entropies. When the groundstate is modified by tuning an external parameter but
remains in the same phase, our best fits to Eq.~(\ref{eq:log_scaling}) (\textit{cf.}
\cite{supplementary})
indicate that $l_q$ varies slightly with the external parameter, even though a universal $l_q$
cannot be excluded. The presence of logarithmic corrections as well as the $l_q$ scalings
$l_{q>1}=q/(q-1) l_\infty$ can be
seen in a toy model of antiferromagnetism as detailed in \cite{supplementary}.

Our set of results on many-body interacting quantum systems shows that subleading terms in the scaling of SR entropies capture universal behavior of quantum states of matter, such as the nature of their broken symmetries or of their criticality. These subleading terms are the natural extensions of fractal dimensions for many-body interacting systems: for instance, we find a $q$-dependence on the first subleading term $b_q$ or $l_q$ only when the system has gapless excitations, analogous to the multifractality at the Anderson transition. 

 QMC being
formulated in terms of path-integrals, a natural extension to
finite-temperature studies is possible.
This could be useful for disordered systems where the many-body localization transition~\cite{MBL}
may be detected in a change of slope of $a_q$ with disorder strength. In another context, one may expect
subleading corrections to capture topological phases 
\cite{moessner_resonating_2001,misguich_quantum_2002,levin_string-net_2005} using a subsystems geometrical
construction~\cite{levin_detecting_2006,kitaev_topological_2006}. Based on previous works on model
wave-functions~\cite{hermanns_renyi_2013,castelnovo_topological_2007} we expect a universal negative topological constant term to show up in SR entropies of any topologically ordered state.

Finally, we emphasize that 
the Monte Carlo schemes presented above are general enough to be used directly with minor
modifications in the study of physical phenomena in different fields, if a stochastic sampling is
applicable~\cite{Note2}
and SR entropies $S_q$ are meaningful.

\begin{acknowledgments}
\makeatletter{}We thank G. Misguich, B. Georgeot, C. Sire  for a critical reading of the manuscript and
very useful suggestions, as well as J.-M. St\'ephan, M. Mambrini, G. Lemari\'e for useful discussions. Our QMC codes
are partly based on the ALPS libraries \cite{ALPS13,ALPS2}. This work was performed using numerical resources from GENCI (grants 2012-x2012050225 and 2013-x2013050225) and CALMIP and is supported by the French ANR program ANR-11-IS04-005-01.

\end{acknowledgments}

\def\url#1{} \def\urlprefix{}\def\url#1{}

\renewenvironment{figure}
{\begin{figure*}}
{\end{figure*}}

\renewenvironment{table}
{\begin{table*}}
{\end{table*}}

\section*{Supplementary material for: \textit{Universal behavior beyond multifractality in quantum
many-body systems}}
{}This supplement contains additional results to strengthen and detail the results presented in the main text of
the report. 

  In Section \ref{sec:2dising_sup},  we discuss the difficulties in demonstrating the convergence of
  the subleading terms $b_q$ towards the thermodynamic limit $-\ln 2$ due to very large entropies
  and thus limited accessible sizes. A study including tilted clusters is presented for the square
  lattice which provides nevertheless convincing evidence for this.
  Additionally, we performed the same calculations for the triangular lattice and give the values of
  the leading and subleading SR entropy scaling coefficients, revealing that the subleading
  terms are \emph{universal} within the same universality class.

  A large number of results for $d=2$ quantum spin systems having a continuous symmetry can be found in Section
  \ref{sec:subleading_sup}, including different variants of the Heisenberg (XXX) and XY models extended by a next
  nearest neighbor ferromagnetic coupling $J_2$ chosen such as to enhance long-range order. Here, we also provide
  information on how exactly fits were performed and provide a statistical measure for the fit
  quality. We also discuss the development of the results towards the thermodynamic limit. The
  argument is completed by an analytically solvable toy model.

  Furthermore, we discuss the technical details of the Monte Carlo method ranging from how to efficiently measure the probabilities $p(\ket{j})$ to the exploitation of symmetries to the
  calculation of entropies from the histogram in Section \ref{sec:practical_sup}.

  For completeness, we note that for all systems studied in this work, the stochastic series
  expansion does not have a sign problem and thus scales polynomially with system size. In fact,
  here the accessible system sizes are rather limited by the linear growth of the entropies, which
  is linked to an exponential decrease of the associated probabilities.

\begin{figure}
\begin{center}
\includegraphics[width=5.5cm]{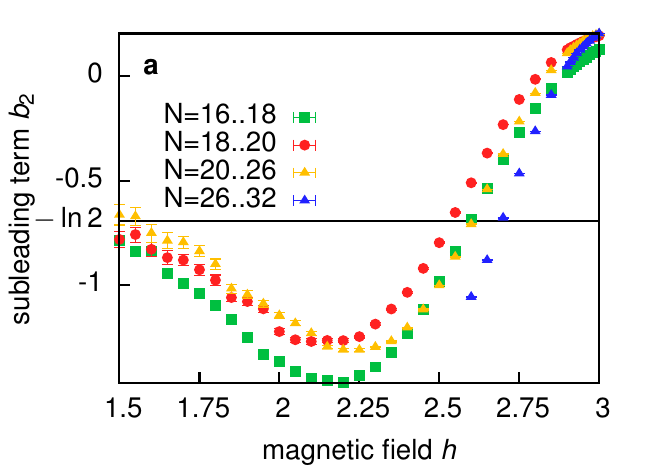}
\includegraphics[width=5.5cm]{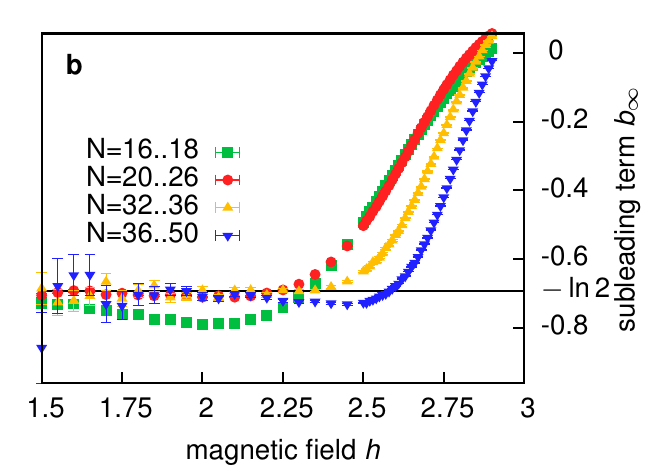}
\end{center}
\caption{Subleading coefficients $b_2$ (a) and $b_\infty$ (b) in the $\sigma^z$ basis for the Ising model on a square
lattice as extracted from small tilted clusters by a fit of $S_q$ to the form $a_q N + b_q$.
The solid line represents the expected thermodynamic value $b_q = -\ln (2)$ below the critical point
$h<h_c\simeq 3.044$. Notice that the non-monotonous convergence to the thermodynamic limit with
cluster size $N$ is typical for tilted clusters with different lattice symmetries.
\label{fig:ising_square_log}}
\end{figure}

\section{2d quantum Ising model in transverse field}
\label{sec:2dising_sup}

\subsection{Square lattice}
\label{subsec:2dising_sup}

The data in the main text in the {\it high-field} polarized phase clearly shows that $b_q$ converges to $0$ in the thermodynamic limit.
In order to support the claim of the universal limit $b_q=-\ln(2)$ in the
{\it low-field} ferromagnetic phase, we perform calculations for small (tilted) square clusters for
$S_2$ and $S_\infty$.
Note that due to the steep behavior of $a_q$ (see for instance panel D of Fig. 2 in the main text for $q=\infty$ -- identical behavior is found for all studied $q$), SR entropies grow very fast with system size in the ferromagnetic phase and do not allow for an accurate calculation for large samples.
Therefore, we perform only a linear fit of the form $S_q=a_q N + b_q$ (see Fig. \ref{fig:ising_square_log}) for different cluster size 
windows of tilted clusters.
Clearly, the oscillation moves closer to the critical point with growing system size and $b_q$
will eventually converge towards the universal value of $-\ln(2)$. 
This is already happening around $N=20$ for low enough fields $h<1.5$ in the case of $b_2$.
Note that a very similar oscillatory finite-size behavior has also been observed close to the
quantum phase transition in the $d=1$ model~\cite{stephan_renyi_2010}. 

The convergence  for $b_\infty$ to the thermodynamic limit 
is much faster, as clearly seen in Fig. \ref{fig:ising_square_log}b.
Moreover, due to smaller $a_\infty$ we can access larger system sizes (see also Fig. 2D in the main
text). This allows us to definitely conclude that the value of $b_q$ in the thermodynamic limit for $h<h_c$ is
$-\ln(2)$.

\begin{figure}
\begin{center}
\includegraphics[width=12cm]{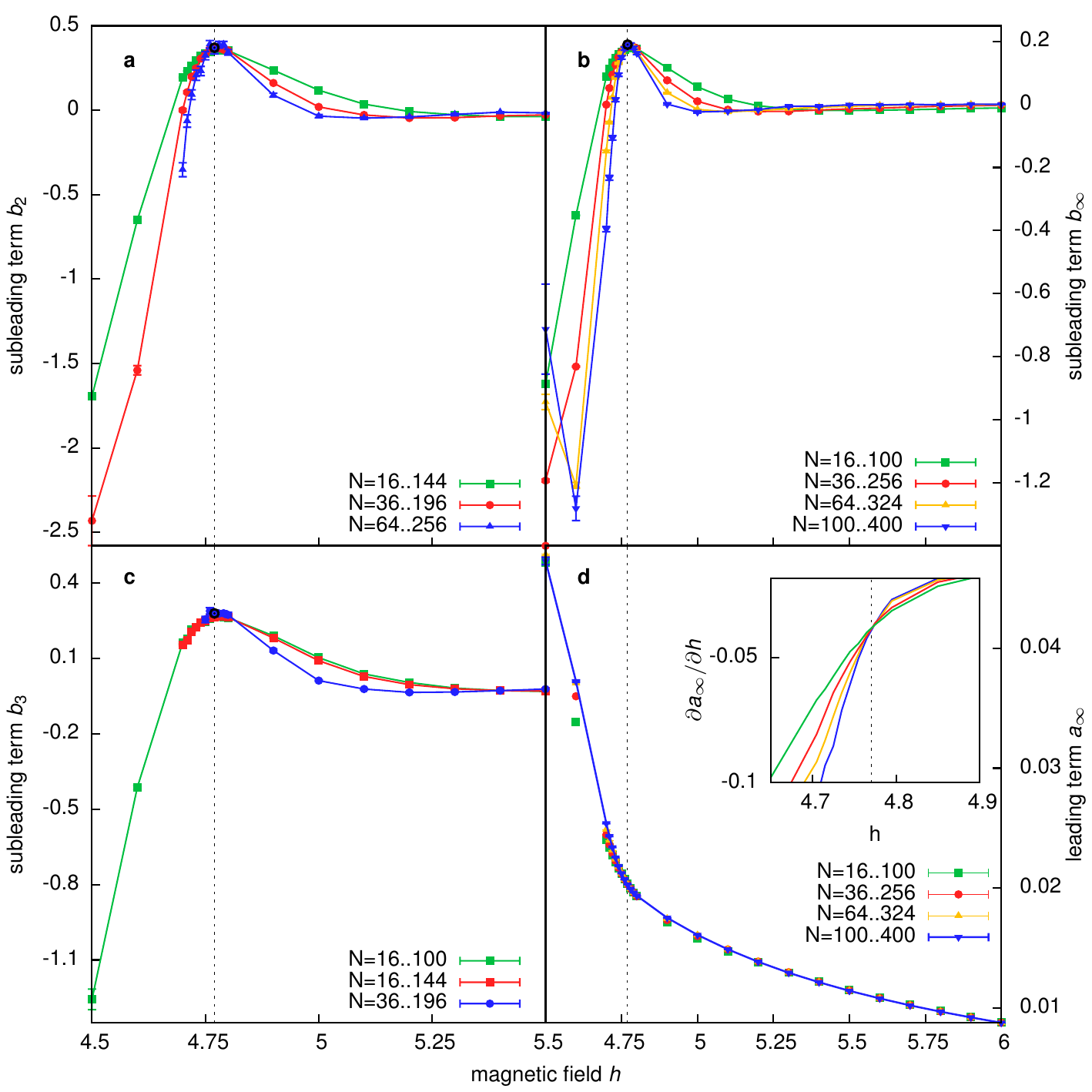}
\end{center}
\caption{\label{fig:ising_triangular} Subleading scaling coefficients $b_q$ in the $\sigma^z$ basis for the Ising model on a triangular
lattice in transverse magnetic field $h$ for different $q$, as obtained from a fit to the form
$S_q=a_q N + b_q + c_q/N + d_q/N^2$.
As in the main text for the square lattice, we show fits over different lattice size
windows in order to demonstrate the development towards the thermodynamic limit. The dashed line
indicates the position of the critical magnetic field $h_c$.
}
\end{figure}

\subsection{Triangular lattice}

\begin{table}
\caption{\label{tab:ising_critical} Leading ($a_q^*$) and subleading ($b_q^*$) scaling coefficient of SR
entropies for the $d=2$ Ising model on the square and triangular lattice at critical magnetic field
$h_c$. Our best estimates indicate that $a^*$ is modified once the lattice is changed, while $b^*$
appears to be universal. Error bars stem from a bootstrap analysis.}
\begin{center}
\begin{tabular}{l|cc}
\hline
\hline
                &   square lattice  &   triangular lattice \\
                & ($h_c\simeq 3.044$) &($h_c\simeq 4.77$)\\
        \hline
$a_2^*$         &    0.04769(2)     &   0.04050(3)                   \\
$a_3^*$         &    0.03588(4)     &   0.03044(6)                   \\
$a_\infty^*$    &    0.02391(3)     &   0.02026(1)                   \\
$b_2^*$         &    0.365(3)       &   0.373(7)                   \\
$b_3^*$         &    0.278(7)       &   0.28(1)                   \\
$b_\infty^*$    &    0.197(25)      &   0.189(4)                   \\
\hline
\hline
\end{tabular}
\end{center}
\end{table}

The universality of the subleading scaling coefficients is demonstrated in a particularly beautiful
way, if we exchange the square lattice by a triangular one.
The position of the critical point
changes ($h_c \simeq 4.77$ for the triangular lattice~\cite{Blote}), leaving the universality class invariant.
Fig.
\ref{fig:ising_triangular} shows the result
for the subleading coefficients $b_2$, $b_3$ and $b_\infty$ in the $\sigma^z$ basis, as well as the leading term $a_\infty$.
While the subleading coefficients in the scaling of the SR entropies adopt \emph{exactly} the same
values (within error bars, cf. Table \ref{tab:ising_critical}) in the thermodynamic limit not only at the critical point but over the \emph{whole range} of magnetic field, the leading term is not identical for different lattices. However, we also find a clear change of slope of $a_q$ (seen best in the derivative displayed in the inset) precisely at the critical point, similar to the square lattice case.

\section{Subleading terms in SR entropy scaling of $d=2$ spin systems with continuous symmetry}
\label{sec:subleading_sup}

\subsection{Best fit results}
For completeness, we report the parameters of best fits of SR entropies for different system sizes
$N\leq 400$ for various models (except for the Heisenberg model (XXX) where the largest accessible
system size has been $N=256$ due to extremely fast growing entropies with $S_\infty>20$ for the
largest size).
Table \ref{tab:best_fit} displays the fit parameters of $S_\infty$ to the form 
\begin{equation}
S_\infty=a_\infty N + l_\infty \ln N + b_\infty + c_\infty/N + d_\infty/N^2,
\end{equation}
 over the \emph{whole} range of system sizes.
In addition to the fit parameters, we report
the fit quality $Q$, which represents the probability of $\chi^2$ assuming its value or being larger,
given the data (including $1\sigma$ error bars) and the number of degrees of freedom as predicted by
the $\chi^2$ distribution (\textit{cf.}
\cite{young_everything_2012}).
Nearly all our results correspond to
outstanding fits signifying that the finite size effects are well captured by $c_q$ and
$d_q$. The data is presented for the Heisenberg isotropic model (XXX), without or with a
ferromagnetic next nearest neighbor couplings $J_2<0$ of the form $J_2 \sum_{<<i,j>>} \vec{S}_i
\cdot \vec{S}_j$, for the XXZ model with easy axis anisotropy $\Delta>1$ as well as for the XY model ($\Delta=0$), which has also been extended by a next
nearest neighbor term $J_2$ of the form $J_2 \sum_{<<i,j>>} S_i^xS_j^x + S_j^yS_j^y$.

\begin{table} 
\caption{\label{tab:best_fit}  Parameters obtained from best fit of the SR entropies (some of which are displayed in Fig. 3. of the main text). Error bars were obtained by a careful bootstrap analysis.
The quality $Q$ (see \cite{young_everything_2012}) of the best fit is indicated in the last column.
Note that they do not reflect by how
much the fit parameters change if one included data for larger system sizes.}
\begin{center}
\begin{tabular}{l|cccccc}
\hline
\hline
model&$a_\infty$ & $l_\infty$ & $b_\infty$ & $c_\infty$ & $d_\infty$ & $Q$\\
\hline
XXX  &  0.0699(3)  &  {0.51(6)}  &  -0.2(2)  &  3(3)  &  -13(16) & 0.45 \\
$\Delta=1.5$ &  0.02585(2)  &  0.001(5)  &  {0.69(2)} &  -0.5(3)  &  25(2) & 0.72 \\
$\Delta=2.0$  &  0.014204(7)  &  -0.003(2)  &  { 0.707(9) }  &  -0.4(1)  &  7.0(8) & 0.74 \\
$J_2=-1.0$  &  0.03371(4)  &  {0.515(8)}  &  0.39(4)  &  -2.0(4)  &  10(3) &
{0.04} \\
$J_2=-5.0$  &  0.01059(8)  &  {0.58(2)}  &  0.75(8)  &  -7.6(10)  &  42(7) & 0.99 \\
XY  &  0.025236(7)  &  {0.269(1)}  &  -0.017(7)  &  -1.40(8)  &  6.1(6) & 0.37 \\
$J_2=-1$  &  0.01498(2)  &  {0.282(3)}  &  0.03(1)  &  -1.7(1)  &  6.1(10) & 0.91 \\
\hline
\hline
\end{tabular}
\end{center}
\end{table}

In Table \ref{tab:Sq}, we show the scaling parameters of SR entropies $S_q$ for different values of
$q$.
The subleading (logarithmic) correction is of particular interest as our results show a clear
dependence of $l_q$ on $q$.
Even though QMC can access much larger system sizes than any other exact
method for the calculation of SR entropies, in some cases the accessible lattices
were still not large enough, mostly due to very fast growing entropies.
Therefore, we present not
only fits including the 5 parameters $a_q$, $l_q$, $b_q$, $c_q$ and $d_q$ but also fits with four
parameters thus leaving out $d_q$ (marked by a dash in the table).
In some cases, we even added a further correction $e_q/N^3$.
If we had only $4$ different system sizes at hand (here, we did only include regular square clusters, {\it i.e.}
no tilted clusters), $Q$ is marked as \emph{n.a.} (not available) as the number of degrees of freedom is
$0$ in this case.
Note that all these fits include the whole range of available system sizes and do
not include an extrapolation to the thermodynamic limit.
Thus, the error bars provided in the table
do not provide information about the bounds of the real value in the thermodynamic limit.

\begin{table}
\caption{\label{tab:Sq}
Parameters from fits to the R\'enyi entropies for different values of $q$.
$4-$, $5-$ and in some cases $6-$ parameter fits are provided for comparison.
}
\begin{center}
\begin{tabular}{l|cccccccc}
\hline
\hline
model      & q        & $a_q$       & $l_q$                       & $b_q$     & $c_q$     & $d_q$  & $e_q$  & $Q$\\
\hline
\hline
XXX	&	&	&	&	&	&	&	&\\
$J_2=0$        & 2        & 0.138(2)    & {1.0(2)}     & -0.8(5)   & 2(2)      & $-$     & $-$ & {n.a.} \\
           & $\infty$ & 0.0699(3)   & {0.51(6)}    & -0.2(2)   & 3(3)      & -13(16) & $-$ & 0.45 \\
           & $\infty$ & 0.07017(7)  & {0.460(5)}   & 0.04(2)   & 0.90(8)   & $-$     & $-$ & 0.51 \\
$J_2=-1$   & $\infty$  &  0.03371(5)  &  0.515(8)  &  0.39(4)  &  -2.0(4)  &  10(3) & $-$ & 0.04 \\
            & $\infty$ &  0.0337(2)  &  0.30(7)  &  1.5(3)  &  -25(6)  &  430(12)  &  -3174(86) & 0.79 \\
$J_2=-5$ & 2        & 0.02013(6)  & {1.373(6)}   & -0.22(2)  & -3.3(1)   & $-$     & $-$ & {0.04} \\
           & 2        & 0.0207(2)   & {1.25(4)}    & 0.3(2)    & -10(2)    & 43(15)  & $-$ & 0.55 \\
           & 3        & 0.0149(4)   & {1.06(3)}    & -0.1(1)   & -1.5(5)   & $-$     & $-$ & 0.97 \\
           & 3        & 0.015(2)    & {1.1(4)}     & -0(2)     & -1(17)    & -4(11)  & $-$ & {n.a.} \\
           & 4        & 0.012(2)    & {1.0(1)}     & -0.3(5)   & 0(2)      & $-$     & $-$ & 0.78 \\
           & 4        & 0.01(1)     & {2(2)}       & -3(9)     & 25(96)    & -153(59)& $-$ & {n.a.}  \\
           & $\infty$ & 0.01010(3)  & {0.689(3)}   & 0.22(1)   & -1.38(6)  & $-$     & $-$ & {0.0} \\
           & $\infty$ & 0.01059(8)  & {0.58(2)}    & 0.75(8)   & -7.6(10)  & 42(7)   & $-$ & 0.99 \\
           & $\infty$ &  0.0107(2)  &  0.55(9)  &  0.9(5)  &  -11(10)  &  104(20)  &  -477(16) & 0.98 \\
           \hline
XY	&	&	&	&	&	&	&	&\\
$J_2=0$         & 2        & 0.04999(6)  & { 0.585(6) } & 0.46(2)   & -0.9(1)   & $-$     & $-$ & 0.1 \\
           & 2        & 0.0505(3)   & {0.50(5)}    & 0.9(2)    & -5(3)     & 27(16)  & $-$ & {n.a.}  \\
           & 3        & 0.0377(2)   & {0.44(2)}    & 0.51(6)   & -0.5(3)   & $-$     & $-$ & 0.7 \\
           & 3        & 0.037(1)    & {0.5(2)}     & 0.1(10)   & 4(10)     & -25(63) & $-$ & {n.a.}  \\
           & 4        & 0.034(1)    & {0.35(8)}    & 0.7(3)    & -1(1)     & $-$     & $-$ & {n.a.} \\
           & $\infty$ & 0.025172(3) & {0.2839(4)}  & -0.087(1) & -0.547(8) & $-$     & $-$ & {0.0} \\
           & $\infty$ & 0.025236(7) & {0.269(2)}   & -0.017(7) & -1.40(9)  & 6.1(6)  & $-$ & 0.37 \\
           & $\infty$ &  0.02520(2)  &  0.281(8)  &  -0.08(4)  &  -0.1(8)  &  -19(16)  &  195(12) & 0.65 \\
$J_2=-1$ & 2        & 0.02975(4)  & {0.598(4)}   & 0.59(2)   & -1.63(8)  & $-$     & $-$ & 0.66 \\
           & 2        & 0.0299(2)   & {0.58(4)}    & 0.7(2)    & -2(2)     & 6(12)   & $-$ & {n.a.} \\
           & 3        & 0.02255(8)  & {0.432(7)}   & 0.67(2)   & -1.4(1)   & $-$     & $-$ & 0.28 \\
           & 3        & 0.0230(4)   & { 0.35(7)}   & 1.0(3)    & -5(3)     & 23(21)  & $-$ & {n.a.} \\
           & 4        & 0.0201(2)   & {0.38(2)}    & 0.70(7)   & -1.4(3)   & $-$     & $-$ & 0.71 \\
           & 4        & 0.021(2)    & { 0.3(3) }   & 1(1)      & -6(12)    & 28(75)  & $-$ & {n.a.} \\
           & $\infty$ & 0.014893(5) & {0.2993(5)}  & -0.053(2) & -0.778(9) & $-$     & $-$ & {0.0} \\
           & $\infty$ & 0.01498(1)  & {0.282(3)}   & 0.03(1)   & -1.7(1)   & 6.1(10) & $-$ & 0.91 \\
           & $\infty$ &  0.01498(5)  &  0.28(1)  &  0.03(7)  &  -2(1)  &  6(23)  &  2(17) & 0.77 \\
\hline
\hline
\end{tabular}
\end{center}
\end{table}

\subsection{Logarithmic corrections in the thermodynamic limit}

In an attempt to extract the value of the logarithmic correction $l_\infty$ in the thermodynamic limit 
 we perform sliding fits of $S_\infty$ to the form 
\begin{equation}
S_\infty= a_\infty N + l_\infty \ln(N) + b_\infty.
\end{equation}
Higher correction terms in $1/N$ will gracefully disappear for large $N$.
The sliding fit is performed such
that from the available data a window of system sizes starting at $N_\mathrm{start}$ of width $5$ is
selected which means \textit{e.g.}
for $N_\mathrm{start}=16$ the system sizes 16, 36, 64, 100 and 144 are
included.
For those system sizes, the fit is performed and the results of the fits are reported in
table \ref{tab:J2_sliding} for the XXX model and in table \ref{tab:J2_sliding_XZ} for the XY
model.
It is obvious that fit qualities are very bad for small system sizes as we do not include
terms $1/N$ in the fit that might correct for finite-size effects.
However, at larger system
sizes (if available), the fit qualities become acceptable and the subleading terms start to converge
to their limiting value.
Certainly, the leading terms $a_\infty$ converge much faster and can be
obtained more easily.

\begin{table}
\caption{
\label{tab:J2_sliding} Parameters obtained from sliding fits of $S_\infty$
for the XXX model with different values of $J_2$.
The fit window size is $5$.
Clearly, the fit quality
becomes only
acceptable for windows comprising larger system sizes as here the $1/N$ and higher correcting terms become irrelevant.}

\begin{center}
XXX, $J_2=0$ \\
\begin{tabular}{l|llll}
\hline
\hline
$N_\mathrm{start}$ & $a_\infty$ & $l_\infty$ & $b_\infty$ & Q \\
\hline
16 & 0.07093(1)&        0.4026(3)&      0.2426(7) & 0.0 \\
36 & 0.07038(4)&        0.430(2)&       0.163(6) & 0.31 \\
64 & 0.0701(2)& 0.45(1)&        0.08(5) & 0.46 \\
\hline
\hline
\end{tabular}\\[1cm]
XXX, $J_2=-1$ \\
\begin{tabular}{l|llll}
\hline
\hline
$N_\mathrm{start}$ & $a_\infty$ & $l_\infty$ & $b_\infty$ & Q \\
\hline
16 & 0.033227(5) &  0.5734(2)&  0.154(4) &0.0 \\
36 & 0.03345(1)  &  0.5597(6)&  0.195(2) & 0.0 \\
64 & 0.03357(2)  &  0.548(2) &  0.237(8) &0.0 \\
100& 0.03374(6)  &  0.523(8) &  0.33(3)  &0.06 \\
144& 0.0334(2)   &  0.59(3)  &  0.04(13) &0.55 \\
\hline
\hline
\end{tabular}\\[1cm]
XXX, $J_2=-5$ \\
\begin{tabular}{l|llll}
\hline
\hline
$N_\mathrm{start}$ & $a_\infty$ & $l_\infty$ & $b_\infty$ & Q \\
\hline
16 & 0.00912(3)  &    0.7713(8) & -0.078(1) & 0.0 \\
36 & 0.00974(3)  &    0.737(1)  & 0.023(6) & 0.0 \\
64 & 0.01015(7)  &    0.690(8)  & 0.19(3) & 0.27 \\
100& 0.01045(12) &    0.644(19) & 0.37(7) & 0.93 \\
144& 0.01044(15) &    0.643(34) & 0.37(15) & 0.91 \\
196& 0.01043(25) &    0.646(69) & 0.36(31) & 0.88 \\
\hline
\hline
\end{tabular}\\[1cm]
\end{center}
\end{table}

\begin{table}
\caption{ \label{tab:J2_sliding_XZ} Sliding fit result for the XY model.
The window size is $5$.
}
\begin{center}
XY, $J_2=0$ \\
\begin{tabular}{l|llll}
\hline
\hline
$N_\mathrm{start}$ & $a_\infty$ & $l_\infty$ & $b_\infty$ & Q \\
\hline
16 & 0.024919(2)&       0.31176(8)&     -0.1938(2) & 0.0 \\
36 & 0.025075(2)&       0.2999(2)&      -0.1555(6) & 0.0 \\
64 & 0.025154(4)&       0.2908(5)&      -0.123(2) & 0.0 \\
100 & 0.025183(6)&      0.2865(10)&     -0.105(4) & 0.63 \\
144 & 0.02518(1)&       0.287(3)&       -0.11(1) & 0.63 \\
\hline
\hline
\end{tabular}\\[1cm]
XY, $J_2=-1$ \\
\begin{tabular}{l|llll}
\hline
\hline
$N_\mathrm{start}$ & $a_\infty$ & $l_\infty$ & $b_\infty$ & Q \\
\hline
16 & 0.014511(1)&       0.33985(7)&     -0.2079(2) & 0.0 \\
36 & 0.014743(3)&       0.3226(2)&      -0.1526(7) & 0.0 \\
64 & 0.014849(7)&       0.3119(6)&      -0.115(2) & 0.01 \\
100 & 0.01492(2)&       0.303(2)&       -0.081(9) & 0.49 \\
\hline
\hline
\end{tabular}
\end{center}
\end{table}

\newpage

\subsection{Shannon-R\'enyi entropies in the Lieb-Mattis model}
\begin{figure}[t]
\begin{center}
\includegraphics[width=3cm]{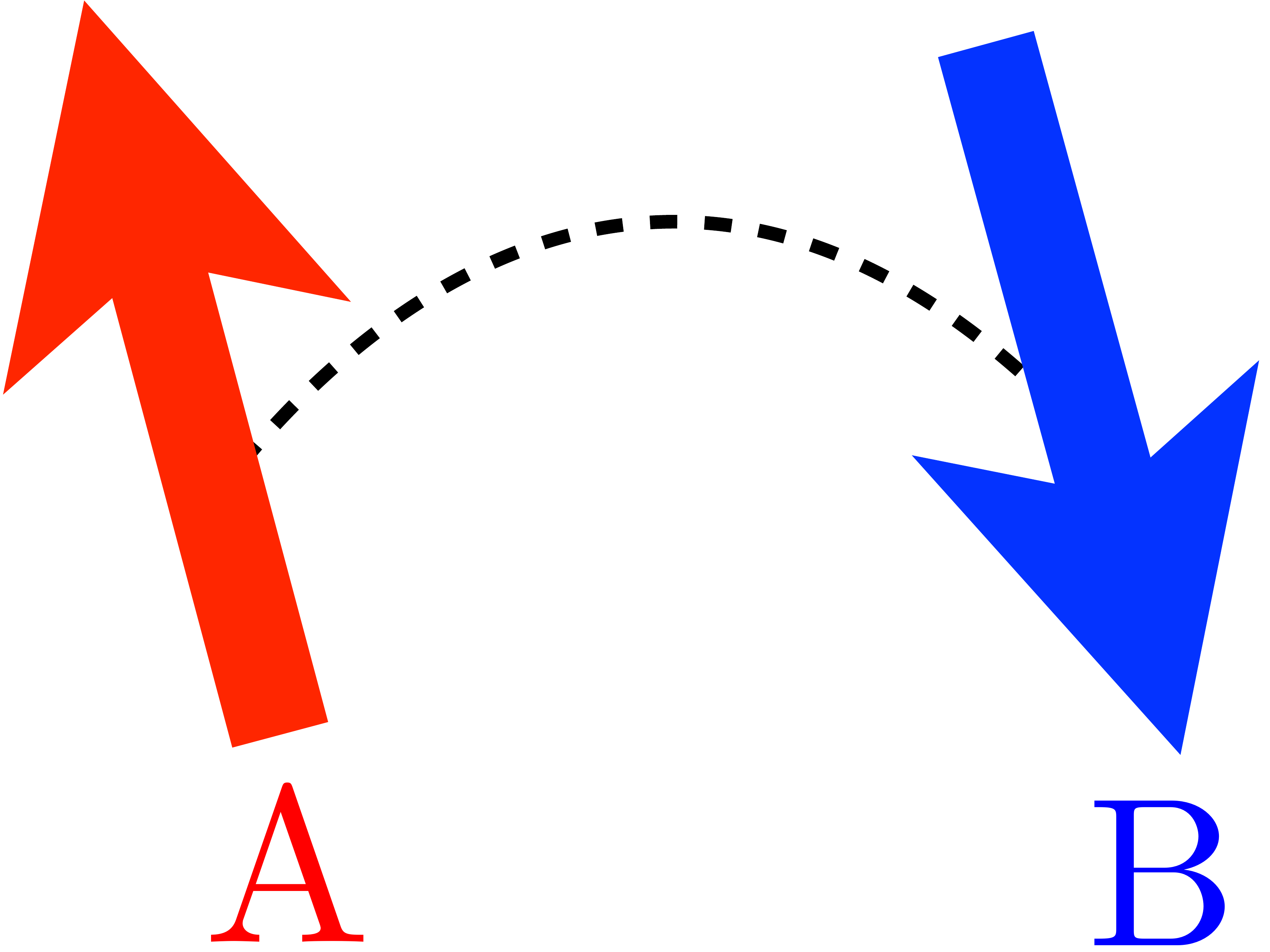}
\end{center}
\caption{Schematic picture for the Lieb-Mattis model of two large spins $S_{\rm A}$ and $S_{\rm B}$ which are antiferromagnetically coupled (dashed line).\label{fig:LM}}
\end{figure}

In order to gain insight onto the logarithmic correction observed in the SR entropies of $d=2$ spin
systems where a continuous symmetry is broken in the thermodynamic limit, one can exactly solve a
simple toy model, introduced in 1962 by Lieb and Mattis in Ref.~\cite{lieb_ordering_1962}.
As depicted in Fig.~\ref{fig:LM}, we consider two large spins ${\bf{S}}_{\rm A}={\bf{S}}_{\rm B}$ of
integer size $N/4$
(each giant spin can be seen as $N/2$ ferromagnetically infinitely coupled microscopic spins $1/2$),
which are coupled antiferromagnetically via
\begin{equation}
H_{\rm LM}={\bf{S}}_{\rm A}\cdot{\bf{S}}_{\rm B}.
\end{equation}
The Hamiltonian conserves the total spins ${\bf S}_A,{\bf S}_B$ and ${\bf S}={\bf S}_A+{\bf S}_B$ and magnetizations $S^z_A,S^z_B$ and $S^z=S^z_A+S^z_B$. The ground-state has total spin $|{\bf S}|=0$, and reads
\begin{equation}
|{\rm{GS}}\rangle=\sum_{M=-N/4}^{N/4}\frac{(-1)^{-M+N/4} }{ \sqrt{1+N/2} }|S^{z}_{\rm A}=M,S^{z}_{\rm B}=-M\rangle.
\end{equation}

Each state $|S^{z}_{\rm A}=M,S^{z}_{\rm B}=-M\rangle$ for fixed $A$ and $B$ magnetizations is the
equal amplitude symmetric state of the microscopic spins $1/2$:  $$|S^{z}_{\rm A}=M,S^{z}_{\rm B}=-M\rangle=\frac{1}{g_M}(\sum_i | i \rangle_A) \otimes (\sum_j | j \rangle_B),$$ with 
\begin{equation}
g_M=\frac{(N/2)!}{(N/4-M)!(N/4+M)!}
\end{equation}
the size of the subspace of magnetization $M$ in $A$ or $B$.

Squaring all amplitudes, the SR entropies for the ground-state of the Lieb-Mattis model for $N$ microscopic spins $1/2$ are given by 
\begin{eqnarray*}
&&S_q(N)=\frac{1}{1-q} \\ 
&&\times \ln\left\{\sum_{i=0}^{N/2}\left[\frac{\frac{N}{2}!}{\left(\frac{N}{2}-i\right)!~i!}\right]^2\times \left(\frac{\left[\left(\frac{N}{2}-i\right)!~i!\right]^2}{\left[\frac{N}{2}!\right]^2\left[1+\frac{N}{2}\right]}\right)^q\right\},
\end{eqnarray*}
which simplifies to
\begin{eqnarray}
S_q(N)&=&\frac{q}{q-1}\ln(1+\frac{N}{2})\nonumber\\ 
&+&\frac{1}{1-q}\ln\left\{\sum_{i=0}^{N/2}\left[\frac{\frac{N}{2}!}{(\frac{N}{2}-i)!i!}\right]^{2(1-q)}\right\}.
\end{eqnarray}
From the above expression, one can immediately notice that $S_\infty=\ln(1+\frac{N}{2})$ and $S_{1/2}=-\ln(1+\frac{N}{2})+{2}\ln(2^{N/2})=N\ln 2-S_\infty$.\\
\\
For $q>1$, one can expand the above sum inside the log:
\begin{eqnarray*}
&&S_q(N)\approx \frac{q}{q-1}\ln(1+\frac{N}{2}) + \frac{\ln 2}{1-q}\\
&&+\frac{1}{1-q}\ln\left(1+\left[\frac{N}{2}\right]^{2-2q}
+\left[\frac{N}{4}\left(\frac{N}{2}-1\right)\right]^{2-2q} + \cdots \right)
\end{eqnarray*}
thus giving 
\begin{equation}
S_q(N)=\frac{q}{q-1}\ln(1+\frac{N}{2})+ \frac{1}{1-q} \ln 2 +  {\cal{O}}(N^{2-2q}),~~~{\rm{if}}~q>1.
\end{equation}
It is interesting to see the that the leading contribution is logarithmic, with no linear term, and that we get 
$$
l_{q>1}=\frac{q}{q-1}l_\infty.
$$
\\
(ii) If $q<1$, the behavior is quite different:
\begin{equation}
S_q(N)=N\ln 2 -\frac{1}{2(1-q)}\ln N +{\cal{O}}(1),~~~{\rm{if}}~q<1.
\end{equation}
(iii) For $q=1$, we get the following behavior for the Shannon entropy:
\begin{equation}
S_1(N)=\frac{N}{2}+1-\ln(2\pi)+{\cal{O}}\left(\frac{\ln N}{N}\right),
\end{equation}
where, surprisingly, the logarithmic corrections have disappeared.
One can also check that for $q=0$, where we expect to recover that $S_0=\ln {\cal D}_0=\ln (N!)-2\ln (N/2!)$, we have indeed, using Vandermonde's identity

\begin{eqnarray*}
S_0&=&\ln \left(\sum_{i=0}^{N/2}\left[\frac{N/2!}{(N/2-i)!i!}\right]^{2}\right)\nonumber\\
&=&\ln \left(\frac{N!}{(N/2!)^2}\right).
\end{eqnarray*}
From this simple Lieb-Mattis toy model for antiferromagnetism, we can see that logarithmic scaling
arises naturally in the behavior of $S_q(N)$ for $q>1$, and are all related to the subleading term
of the most probable state at $q=\infty$ by $l_{q>1}=\frac{q}{(q-1)}l_\infty$. It is interesting to
note that the most probable state $|\Uparrow_{\rm{A}}\Downarrow_{\rm B}\rangle =
|\uparrow\cdots\uparrow\rangle_{\rm A}\otimes |\downarrow\cdots\downarrow\rangle_{\rm B}$ dominates
the groundstate with a probability $p_{\rm max}=1/(1+N/2)$, while the second most probable state has
$p'_{\rm max}=4/(N^2+N^3/2)$ (with a multiplicity $N^2/4$), meaning that the first gap in the
spectrum $\ln p_{\rm max}-\ln p'_{\rm max}$ grows as $\ln N$ when $N$ increases.

We finally remark that the value $l_\infty=1$ that we obtain cannot be directly compared with the
results on the $SU(2)$  Heisenberg model. Indeed, while the Lieb-Mattis model correctly captures the
Anderson tower of states~\cite{Anderson} needed for $SU(2)$ symmetry breaking, it does {\it not} exhibit gapless Goldstone modes (see {\it e.g.} Ref.~\cite{lhuillier_frustrated_2005}). Another artifact due to the infinite-range of interactions in the model are the peculiar values $a_{0 \leq q < 1} = \ln {2} , a_1=1/2, a_{q>1}=0$ ({\it i.e.} no leading linear term for $q>1$) in the scaling of SR entropies, which are clearly not generic.

\section{Practical issues for measuring SR entropies with quantum Monte Carlo}
\label{sec:practical_sup}

\subsection{Collecting measurements for $p(|j\rangle)$}

Measuring $p(|j\rangle)$ for all configurations $|j\rangle$ encountered in QMC can rapidly become a
technical bottleneck as the amount of required statistics is very high.
If it fits in memory, the probably simpler solution is to record the occurrences of state $|j\rangle$ in a histogram $H(|j\rangle)$ in a pre-declared large array, which index is given by the bit representation of $|j\rangle$ (for $N\leq 64$ spins $1/2$, the representation of a state fits in a 64-bit integer).
When memory issues become problematic, a more judicious solution is provided by using associate
arrays (maps), with the key being the state bit representation of $|j\rangle$ and the value
$H(|j\rangle)$. In memory sparse situations, we found it advantageous to use
\texttt{google::sparse\_hash\_map} \cite{google_sparsehash} which is an associative array that uses a particularly high level of memory efficiency at the
expense of computer time.
Each time a state $|j\rangle$ is observed, one can check whether it is already present in the map (in which case the corresponding value is incremented by one) or not (in which case a new key is added to the map).
In this way, only states that are actually observed are stored.

This is often not enough.
In particular, we find that it is not possible to store all probabilities in memory for a QMC run for the largest systems (typically for $N>36$ for the transverse-field Ising model, $N>40$ for the XXZ model).
Note that this crucially depends on the SR entropy itself (the larger $S_q$, the more probabilities need to be stored).
In the case where the memory limit is reached, we dump the histogram to disk, and reset the map in memory before continuing the QMC simulation.
Once the simulation is finished, we add all histograms dumped into disk into a single histogram.
In order to add histograms (and to again avoid memory issues), we find very useful to save states to disk using the natural order of the state bit representation, such that one can directly write to disk (without storing in memory) the new histogram obtained by adding two histograms.
Once the final histogram is written to disk, entropies can be computed (see below) without reloading the full histogram in memory.

As memory is an issue, it is quite crucial to furthermore use symmetries of the problem in the implementation of histograms in order to reduce their sizes.

\subsection{Use of symmetries}

{\it Choice of computational basis --- } If in the QMC simulation there is freedom in the choice of
the computational basis, it is judicious to consider the basis that is computationally feasible and where the ground-state wave-function
has the lowest entropy. For instance, for the transverse-field Ising model, it is possible to perform simulations in the $\{\sigma^x\}$~\cite{Sandvik03} or $\{\sigma^z\}$~\cite{Albuquerque10} basis, allowing to measure entropies $S_q^x$ and $S_q^z$ in the different basis.
In one dimension, the two entropies are related by duality $S_q^x(h)=S_q^z(1/h)+\ln (2)$
\cite{stephan_renyi_2010}, and one can therefore choose the basis in which $S_q$ is the lowest for
the specific values of $h$ simulated.
Note that at the critical point $h_c=1$ in one dimension, there is not much to be gained by changing
the basis.
In any dimension, it is easy to show that the following relation holds for all $h$:
\begin{equation}
    S_{\frac{1}{2}}^{x,z} = N \ln 2 - S_{\infty}^{z,x}.
\end{equation}
We observe that in most of the phase diagram (and especially close to the critical point $h_c\simeq
3.044$) for the square lattice, the SR entropies $S_q^z$ in the $\{\sigma^z\}$ basis are much lower
than in the $\{\sigma^x\}$ basis -- this explains why our results are shown in this basis.
The same argument applies for the $d=2$ XY model, which can be simulated both in $\{S^z\}$ and $\{S^x\}$ basis~\cite{Sandvik99}, the latter turning out to have lower SR entropies, as can easily be understood from the nature of the ground-state.
For the XXX Heisenberg model, SR entropies $S_q$ are obviously identical in the $\{S^x\}$, $\{S^y\}$
and $\{S^z\}$ bases as well as in all other bases linked by global SU(2) transformations.

{\it  Conserved quantum numbers of the Hamiltonian --- } If the ground-state wave-function belongs to a particular symmetry sector of the Hamiltonian, it is useful to only perform measurements of $p(|j\rangle)$ in this sector.
For instance, the ground-state of the transverse-field Ising model (with PBC) belongs to the parity
sector $P=1$, \textit{i.e.} the eigenvalue 1 of the parity operator $\hat{P}=\prod_{i=1}^{N} \sigma_i^z$.
When observed, a state $|j\rangle$ is stored in the histogram only if it belongs to the sector $P=1$.
Note that one needs to be able to easily check the symmetry sector of a state, which means that the symmetry operator needs to be {\it diagonal} in the computational basis.
This is the case for $\hat{P}$ in the transverse-field Ising model when the  $\{\sigma^z\}$ basis is used.
For the XXZ model, one can easily compute the magnetization $S^z$ of the state, which is also a conserved quantity diagonal in the $\{S^z\}$ basis.

This use of symmetry is useful for two reasons : (i) the temperature where the ground-state is reached is {\it higher} when the measurements are restricted to the ground-state symmetry sector: indeed the gap to the first excited state in the same sector is usually {\it larger} than the gap to the first excited state in another sector,
(ii) less states need to be stored in the histogram, reducing its size.
For the transverse-field Ising model, this corresponds to a gain of a factor $2$, and of a factor up to  $N! / [ (N/2)! ]^2$ for the XXZ model which is considerable.

{\it Space group symmetries --- } States are also invariant under lattice symmetries.
The use of the latter is usually more involved.
The ground-state of interest is in most cases located in the most symmetric  sector, which provides a huge gain in histogram size.
We use the following technique to implement this symmetry: every time a state $|j\rangle$ is encountered, we find its {\it parent}  ${\cal P}(|j\rangle)$, {\it i.e.} the state with the lowest bit representation which is equivalent to $|j\rangle$ by symmetry.
To find ${\cal P}(|j\rangle )$, we generate a set of new states by applying all lattice symmetries to $|j\rangle$, and find in this set the state with the lowest bit representation.
We also store the degeneracy ${\rm deg}({\cal P}(|j\rangle ))$ of the set (the number of independent states generated in the process) on disk as this will be needed for the final calculation of entropies.
Note that we find that storing the parent or its degeneracy for every state is too demanding in terms of memory, and we find it more efficient to recompute it each time.

For the square lattices with total number of sites $N$, the total number of symmetries is $8 N$ (for regular clusters of size $N=L^2$ , or tilted clusters of size $N=p^2+p^2$ with $p$ integer) or $4N$ (for other tilted clusters of shape $N=m^2+n^2$ with $m,n$ different integers), where the factor $N$ comes from translations and the factor $4$ or $8$ from independent rotations~\cite{sandvik_computational_2010}.
We find that the degeneracy ${\rm deg}({\cal P}(|j\rangle))$ is precisely equal to the number of symmetries of the cluster for most states $|j\rangle$, therefore we only save to disk the degeneracies ${\rm deg}({\cal P}(|j\rangle ))$ when this is {\it not} the case.

Imposing the space group symmetries improves very much the convergence of $S_q$ as not only much less memory is needed to store the histogram, but also much less measurements are needed (in some sense, when one encounters a state $|j\rangle$, all symmetry-equivalent states are considered to be encountered, which speeds up the simulation).

{\it Imaginary time translation --- } Since all time slices are equivalent in QMC (cyclicity of the trace in the partition function), one can improve statistics by measuring $p(|j\rangle)$ at any time step in path-integral QMC (or propagation index in Stochastic Series Expansion).
Statistics are improved by a factor $\beta N$, which also means that one needs to be able to save $\beta N$ more measurements.

\subsection{Reconstructing entropies from histogram}

Once the final histogram $H({\cal P}(|j\rangle)$ as well as the degeneracy ${\rm deg}({\cal P}(|j\rangle))$  are obtained for the parent states, we first compute intermediate sums 
\begin{eqnarray}
\tilde{Z} & = & \sum_{\cal P} H({\cal P}) \\
\tilde{S_1} & = & -\sum_{\cal P} H({\cal P}).\ln[H({\cal P})/{\rm deg}({\cal P})] \\
 \tilde{S_q} & = & \sum_{\cal P} [ H({\cal P}) / {\rm deg}({\cal P}) ]^q.
{\rm deg}({\cal P})
\end{eqnarray}
to finally obtain the SR entropies:
\begin{eqnarray}
S_1 & = & \tilde{S_1}/\tilde{Z}+\ln{\tilde{Z}} \\
S_q & = & \frac{1}{1-q} [ \ln{\tilde{S_q}} - q  \ln{\tilde{Z}} ].
\end{eqnarray}

\def\url#1{} \def\urlprefix{}\def\url#1{}


\begin{thebibliography}{46}\makeatletter
\providecommand \@ifxundefined [1]{ \@ifx{#1\undefined}
}\providecommand \@ifnum [1]{ \ifnum #1\expandafter \@firstoftwo
 \else \expandafter \@secondoftwo
 \fi
}\providecommand \@ifx [1]{ \ifx #1\expandafter \@firstoftwo
 \else \expandafter \@secondoftwo
 \fi
}\providecommand \natexlab [1]{#1}\providecommand \enquote  [1]{``#1''}\providecommand \bibnamefont  [1]{#1}\providecommand \bibfnamefont [1]{#1}\providecommand \citenamefont [1]{#1}\providecommand \href@noop [0]{\@secondoftwo}\providecommand \href [0]{\begingroup \@sanitize@url \@href}\providecommand \@href[1]{\@@startlink{#1}\@@href}\providecommand \@@href[1]{\endgroup#1\@@endlink}\providecommand \@sanitize@url [0]{\catcode `\\12\catcode `\$12\catcode
  `\&12\catcode `\#12\catcode `\^12\catcode `\_12\catcode `\%12\relax}\providecommand \@@startlink[1]{}\providecommand \@@endlink[0]{}\providecommand \url  [0]{\begingroup\@sanitize@url \@url }\providecommand \@url [1]{\endgroup\@href {#1}{\urlprefix }}\providecommand \urlprefix  [0]{URL }\providecommand \Eprint [0]{\href }\providecommand \doibase [0]{http://dx.doi.org/}\providecommand \selectlanguage [0]{\@gobble}\providecommand \bibinfo  [0]{\@secondoftwo}\providecommand \bibfield  [0]{\@secondoftwo}\providecommand \translation [1]{[#1]}\providecommand \BibitemOpen [0]{}\providecommand \bibitemStop [0]{}\providecommand \bibitemNoStop [0]{.\EOS\space}\providecommand \EOS [0]{\spacefactor3000\relax}\providecommand \BibitemShut  [1]{\csname bibitem#1\endcsname}\let\auto@bib@innerbib\@empty
\bibitem [{\citenamefont {Evers}\ and\ \citenamefont
  {Mirlin}(2008)}]{evers_anderson_2008}  \BibitemOpen
  \bibfield  {author} {\bibinfo {author} {\bibfnamefont {F.}~\bibnamefont
  {Evers}}\ and\ \bibinfo {author} {\bibfnamefont {A.~D.}\ \bibnamefont
  {Mirlin}},\ }\href {\doibase 10.1103/RevModPhys.80.1355} {\bibfield
  {journal} {\bibinfo  {journal} {Reviews of Modern Physics}\ }\textbf
  {\bibinfo {volume} {80}},\ \bibinfo {pages} {1355} (\bibinfo {year}
  {2008})}\BibitemShut {NoStop}\bibitem [{\citenamefont {Evers}\ and\ \citenamefont
  {Mirlin}(2000)}]{evers_fluctuations_2000}  \BibitemOpen
  \bibfield  {author} {\bibinfo {author} {\bibfnamefont {F.}~\bibnamefont
  {Evers}}\ and\ \bibinfo {author} {\bibfnamefont {A.}~\bibnamefont {Mirlin}},\
  }\href {\doibase 10.1103/PhysRevLett.84.3690} {\bibfield  {journal} {\bibinfo
   {journal} {Physical Review Letters}\ }\textbf {\bibinfo {volume} {84}},\
  \bibinfo {pages} {3690} (\bibinfo {year} {2000})}\BibitemShut {NoStop}\bibitem [{\citenamefont {Grassberger}(1986)}]{grassberger_complexity_1986}  \BibitemOpen
  \bibfield  {author} {\bibinfo {author} {\bibfnamefont {P.}~\bibnamefont
  {Grassberger}},\ }\href {\doibase 10.1007/BF00668821} {\bibfield  {journal}
  {\bibinfo  {journal} {International Journal of Theoretical Physics}\ }\textbf
  {\bibinfo {volume} {25}},\ \bibinfo {pages} {907} (\bibinfo {year}
  {1986})}\BibitemShut {NoStop}\bibitem [{\citenamefont {Evangelou}\ and\ \citenamefont
  {Pichard}(2000)}]{evangelou_2000}  \BibitemOpen
  \bibfield  {author} {\bibinfo {author} {\bibfnamefont {S.~N.}\ \bibnamefont
  {Evangelou}}\ and\ \bibinfo {author} {\bibfnamefont {J.-L.}\ \bibnamefont
  {Pichard}},\ }\href {\doibase 10.1103/PhysRevLett.84.1643} {\bibfield
  {journal} {\bibinfo  {journal} {Physical Review Letters}\ }\textbf {\bibinfo
  {volume} {84}},\ \bibinfo {pages} {1643} (\bibinfo {year}
  {2000})}\BibitemShut {NoStop}\bibitem [{\citenamefont {Stanley}\ and\ \citenamefont
  {Meakin}(1988)}]{stanley_multifractal_1988}  \BibitemOpen
  \bibfield  {author} {\bibinfo {author} {\bibfnamefont {H.~E.}\ \bibnamefont
  {Stanley}}\ and\ \bibinfo {author} {\bibfnamefont {P.}~\bibnamefont
  {Meakin}},\ }\href {\doibase 10.1038/335405a0} {\bibfield  {journal}
  {\bibinfo  {journal} {Nature}\ }\textbf {\bibinfo {volume} {335}},\ \bibinfo
  {pages} {405} (\bibinfo {year} {1988})}\BibitemShut {NoStop}\bibitem [{\citenamefont {Georgeot}\ and\ \citenamefont
  {Shepelyansky}(2000)}]{georgeot_quantum_2000}  \BibitemOpen
  \bibfield  {author} {\bibinfo {author} {\bibfnamefont {B.}~\bibnamefont
  {Georgeot}}\ and\ \bibinfo {author} {\bibfnamefont {D.}~\bibnamefont
  {Shepelyansky}},\ }\href {\doibase 10.1103/PhysRevE.62.3504} {\bibfield
  {journal} {\bibinfo  {journal} {Physical Review E}\ }\textbf {\bibinfo
  {volume} {62}},\ \bibinfo {pages} {3504} (\bibinfo {year}
  {2000})}\BibitemShut {NoStop}\bibitem [{\citenamefont {Halsey}\ \emph {et~al.}(1986)\citenamefont {Halsey},
  \citenamefont {Jensen}, \citenamefont {Kadanoff}, \citenamefont {Procaccia},\
  and\ \citenamefont {Shraiman}}]{halsey_multifractal_1986}  \BibitemOpen
  \bibfield  {author} {\bibinfo {author} {\bibfnamefont {T.~C.}\ \bibnamefont
  {Halsey}}, \bibinfo {author} {\bibfnamefont {M.~H.}\ \bibnamefont {Jensen}},
  \bibinfo {author} {\bibfnamefont {L.~P.}\ \bibnamefont {Kadanoff}}, \bibinfo
  {author} {\bibfnamefont {I.}~\bibnamefont {Procaccia}}, \ and\ \bibinfo
  {author} {\bibfnamefont {B.~I.}\ \bibnamefont {Shraiman}},\ }\href {\doibase
  10.1103/PhysRevA.33.1141} {\bibfield  {journal} {\bibinfo  {journal}
  {Physical Review A}\ }\textbf {\bibinfo {volume} {33}},\ \bibinfo {pages}
  {1141} (\bibinfo {year} {1986})}\BibitemShut {NoStop}\bibitem [{Note1()}]{Note1}  \BibitemOpen
  \bibinfo {note} {These entropies are closely related to the generalized
  inverse participation ratios $\DOTSB \sum@ \slimits@ _i p_i^{q}$, which are
  well-studied objects for instance in Anderson localization~\cite
  {evers_anderson_2008,evers_fluctuations_2000}.}\BibitemShut {Stop}\bibitem [{\citenamefont {Janssen}(1994)}]{janssen_multifractal_1994}  \BibitemOpen
  \bibfield  {author} {\bibinfo {author} {\bibfnamefont {M.}~\bibnamefont
  {Janssen}},\ }\href {\doibase 10.1142/S021797929400049X} {\bibfield
  {journal} {\bibinfo  {journal} {International Journal of Modern Physics B}\
  }\textbf {\bibinfo {volume} {08}},\ \bibinfo {pages} {943} (\bibinfo {year}
  {1994})}\BibitemShut {NoStop}\bibitem [{Note2()}]{Note2}  \BibitemOpen
  \bibinfo {note} {These methods are efficient when the underlying QMC is,
  {\protect \it i.e.} for models with no sign problem.}\BibitemShut {Stop}\bibitem [{\citenamefont {Atas}\ and\ \citenamefont
  {Bogomolny}(2012)}]{atas_multifractality_2012}  \BibitemOpen
  \bibfield  {author} {\bibinfo {author} {\bibfnamefont {Y.~Y.}\ \bibnamefont
  {Atas}}\ and\ \bibinfo {author} {\bibfnamefont {E.}~\bibnamefont
  {Bogomolny}},\ }\href {\doibase 10.1103/PhysRevE.86.021104} {\bibfield
  {journal} {\bibinfo  {journal} {Physical Review E}\ }\textbf {\bibinfo
  {volume} {86}},\ \bibinfo {pages} {021104} (\bibinfo {year}
  {2012})}\BibitemShut {NoStop}\bibitem [{\citenamefont {St{\'e}phan}\ \emph {et~al.}(2009)\citenamefont
  {St{\'e}phan}, \citenamefont {Furukawa}, \citenamefont {Misguich},\ and\
  \citenamefont {Pasquier}}]{stephan_shannon_2009}  \BibitemOpen
  \bibfield  {author} {\bibinfo {author} {\bibfnamefont {J.-M.}\ \bibnamefont
  {St{\'e}phan}}, \bibinfo {author} {\bibfnamefont {S.}~\bibnamefont
  {Furukawa}}, \bibinfo {author} {\bibfnamefont {G.}~\bibnamefont {Misguich}},
  \ and\ \bibinfo {author} {\bibfnamefont {V.}~\bibnamefont {Pasquier}},\
  }\href {\doibase 10.1103/PhysRevB.80.184421} {\bibfield  {journal} {\bibinfo
  {journal} {Physical Review B}\ }\textbf {\bibinfo {volume} {80}},\ \bibinfo
  {pages} {184421} (\bibinfo {year} {2009})}\BibitemShut {NoStop}\bibitem [{\citenamefont {St{\'e}phan}\ \emph {et~al.}(2010)\citenamefont
  {St{\'e}phan}, \citenamefont {Misguich},\ and\ \citenamefont
  {Pasquier}}]{stephan_renyi_2010}  \BibitemOpen
  \bibfield  {author} {\bibinfo {author} {\bibfnamefont {J.-M.}\ \bibnamefont
  {St{\'e}phan}}, \bibinfo {author} {\bibfnamefont {G.}~\bibnamefont
  {Misguich}}, \ and\ \bibinfo {author} {\bibfnamefont {V.}~\bibnamefont
  {Pasquier}},\ }\href {\doibase 10.1103/PhysRevB.82.125455} {\bibfield
  {journal} {\bibinfo  {journal} {Physical Review B}\ }\textbf {\bibinfo
  {volume} {82}},\ \bibinfo {pages} {125455} (\bibinfo {year}
  {2010})}\BibitemShut {NoStop}\bibitem [{\citenamefont {St{\'e}phan}\ \emph {et~al.}(2011)\citenamefont
  {St{\'e}phan}, \citenamefont {Misguich},\ and\ \citenamefont
  {Pasquier}}]{stephan_phase_2011}  \BibitemOpen
  \bibfield  {author} {\bibinfo {author} {\bibfnamefont {J.-M.}\ \bibnamefont
  {St{\'e}phan}}, \bibinfo {author} {\bibfnamefont {G.}~\bibnamefont
  {Misguich}}, \ and\ \bibinfo {author} {\bibfnamefont {V.}~\bibnamefont
  {Pasquier}},\ }\href {\doibase 10.1103/PhysRevB.84.195128} {\bibfield
  {journal} {\bibinfo  {journal} {Physical Review B}\ }\textbf {\bibinfo
  {volume} {84}},\ \bibinfo {pages} {195128} (\bibinfo {year}
  {2011})}\BibitemShut {NoStop}\bibitem [{\citenamefont {Alcaraz}\ and\ \citenamefont
  {Rajabpour}(2013)}]{alcaraz_universal_2013}  \BibitemOpen
  \bibfield  {author} {\bibinfo {author} {\bibfnamefont {F.~C.}\ \bibnamefont
  {Alcaraz}}\ and\ \bibinfo {author} {\bibfnamefont {M.~A.}\ \bibnamefont
  {Rajabpour}},\ }\href {\doibase 10.1103/PhysRevLett.111.017201} {\bibfield
  {journal} {\bibinfo  {journal} {Physical Review Letters}\ }\textbf {\bibinfo
  {volume} {111}},\ \bibinfo {pages} {017201} (\bibinfo {year}
  {2013})}\BibitemShut {NoStop}\bibitem [{\citenamefont {Amico}\ \emph {et~al.}(2008)\citenamefont {Amico},
  \citenamefont {Fazio}, \citenamefont {Osterloh},\ and\ \citenamefont
  {Vedral}}]{amico_entanglement_2008}  \BibitemOpen
  \bibfield  {author} {\bibinfo {author} {\bibfnamefont {L.}~\bibnamefont
  {Amico}}, \bibinfo {author} {\bibfnamefont {R.}~\bibnamefont {Fazio}},
  \bibinfo {author} {\bibfnamefont {A.}~\bibnamefont {Osterloh}}, \ and\
  \bibinfo {author} {\bibfnamefont {V.}~\bibnamefont {Vedral}},\ }\href
  {\doibase 10.1103/RevModPhys.80.517} {\bibfield  {journal} {\bibinfo
  {journal} {Reviews of Modern Physics}\ }\textbf {\bibinfo {volume} {80}},\
  \bibinfo {pages} {517} (\bibinfo {year} {2008})}\BibitemShut {NoStop}\bibitem [{\citenamefont {Calabrese}\ and\ \citenamefont
  {Cardy}(2004)}]{calabrese_entanglement_2004}  \BibitemOpen
  \bibfield  {author} {\bibinfo {author} {\bibfnamefont {P.}~\bibnamefont
  {Calabrese}}\ and\ \bibinfo {author} {\bibfnamefont {J.}~\bibnamefont
  {Cardy}},\ }\href {\doibase 10.1088/1742-5468/2004/06/P06002} {\bibfield
  {journal} {\bibinfo  {journal} {Journal of Statistical Mechanics: Theory and
  Experiment}\ }\textbf {\bibinfo {volume} {2004}},\ \bibinfo {pages} {P06002}
  (\bibinfo {year} {2004})}\BibitemShut {NoStop}\bibitem [{\citenamefont {Levin}\ and\ \citenamefont
  {Wen}(2006)}]{levin_detecting_2006}  \BibitemOpen
  \bibfield  {author} {\bibinfo {author} {\bibfnamefont {M.}~\bibnamefont
  {Levin}}\ and\ \bibinfo {author} {\bibfnamefont {X.-G.}\ \bibnamefont
  {Wen}},\ }\href {\doibase 10.1103/PhysRevLett.96.110405} {\bibfield
  {journal} {\bibinfo  {journal} {Physical Review Letters}\ }\textbf {\bibinfo
  {volume} {96}},\ \bibinfo {pages} {110405} (\bibinfo {year}
  {2006})}\BibitemShut {NoStop}\bibitem [{\citenamefont {Kitaev}\ and\ \citenamefont
  {Preskill}(2006)}]{kitaev_topological_2006}  \BibitemOpen
  \bibfield  {author} {\bibinfo {author} {\bibfnamefont {A.}~\bibnamefont
  {Kitaev}}\ and\ \bibinfo {author} {\bibfnamefont {J.}~\bibnamefont
  {Preskill}},\ }\href {\doibase 10.1103/PhysRevLett.96.110404} {\bibfield
  {journal} {\bibinfo  {journal} {Physical Review Letters}\ }\textbf {\bibinfo
  {volume} {96}},\ \bibinfo {pages} {110404} (\bibinfo {year}
  {2006})}\BibitemShut {NoStop}\bibitem [{\citenamefont {Isakov}\ \emph {et~al.}(2011)\citenamefont {Isakov},
  \citenamefont {Hastings},\ and\ \citenamefont
  {Melko}}]{isakov_topological_2011}  \BibitemOpen
  \bibfield  {author} {\bibinfo {author} {\bibfnamefont {S.~V.}\ \bibnamefont
  {Isakov}}, \bibinfo {author} {\bibfnamefont {M.~B.}\ \bibnamefont
  {Hastings}}, \ and\ \bibinfo {author} {\bibfnamefont {R.~G.}\ \bibnamefont
  {Melko}},\ }\href {\doibase 10.1038/nphys2036} {\bibfield  {journal}
  {\bibinfo  {journal} {Nature Physics}\ }\textbf {\bibinfo {volume} {7}},\
  \bibinfo {pages} {772} (\bibinfo {year} {2011})}\BibitemShut {NoStop}\bibitem [{\citenamefont {Jiang}\ \emph {et~al.}(2012)\citenamefont {Jiang},
  \citenamefont {Wang},\ and\ \citenamefont
  {Balents}}]{jiang_identifying_2012}  \BibitemOpen
  \bibfield  {author} {\bibinfo {author} {\bibfnamefont {H.-C.}\ \bibnamefont
  {Jiang}}, \bibinfo {author} {\bibfnamefont {Z.}~\bibnamefont {Wang}}, \ and\
  \bibinfo {author} {\bibfnamefont {L.}~\bibnamefont {Balents}},\ }\href
  {\doibase 10.1038/nphys2465} {\bibfield  {journal} {\bibinfo  {journal}
  {Nature Physics}\ }\textbf {\bibinfo {volume} {8}},\ \bibinfo {pages} {902}
  (\bibinfo {year} {2012})}\BibitemShut {NoStop}\bibitem [{\citenamefont {Metlitski}\ and\ \citenamefont
  {Grover}(2011)}]{metlitski_entanglement_2011}  \BibitemOpen
  \bibfield  {author} {\bibinfo {author} {\bibfnamefont {M.~A.}\ \bibnamefont
  {Metlitski}}\ and\ \bibinfo {author} {\bibfnamefont {T.}~\bibnamefont
  {Grover}},\ }\href {http://arxiv.org/abs/1112.5166} {\emph {\bibinfo {title}
  {Entanglement Entropy of Systems with Spontaneously Broken Continuous
  Symmetry}}},\ \bibinfo {type} {{arXiv} e-print}\ \bibinfo {number}
  {1112.5166}\ (\bibinfo {year} {2011})\BibitemShut {NoStop}\bibitem [{\citenamefont {Kallin}\ \emph {et~al.}(2011)\citenamefont {Kallin},
  \citenamefont {Hastings}, \citenamefont {Melko},\ and\ \citenamefont
  {Singh}}]{kallin_anomalies_2011}  \BibitemOpen
  \bibfield  {author} {\bibinfo {author} {\bibfnamefont {A.~B.}\ \bibnamefont
  {Kallin}}, \bibinfo {author} {\bibfnamefont {M.~B.}\ \bibnamefont
  {Hastings}}, \bibinfo {author} {\bibfnamefont {R.~G.}\ \bibnamefont {Melko}},
  \ and\ \bibinfo {author} {\bibfnamefont {R.~R.~P.}\ \bibnamefont {Singh}},\
  }\href {\doibase 10.1103/PhysRevB.84.165134} {\bibfield  {journal} {\bibinfo
  {journal} {Physical Review B}\ }\textbf {\bibinfo {volume} {84}},\ \bibinfo
  {pages} {165134} (\bibinfo {year} {2011})}\BibitemShut {NoStop}\bibitem [{\citenamefont {Kallin}\ \emph {et~al.}(2013)\citenamefont {Kallin},
  \citenamefont {Hyatt}, \citenamefont {Singh},\ and\ \citenamefont
  {Melko}}]{kallin_entanglement_2013}  \BibitemOpen
  \bibfield  {author} {\bibinfo {author} {\bibfnamefont {A.~B.}\ \bibnamefont
  {Kallin}}, \bibinfo {author} {\bibfnamefont {K.}~\bibnamefont {Hyatt}},
  \bibinfo {author} {\bibfnamefont {R.~R.~P.}\ \bibnamefont {Singh}}, \ and\
  \bibinfo {author} {\bibfnamefont {R.~G.}\ \bibnamefont {Melko}},\ }\href
  {\doibase 10.1103/PhysRevLett.110.135702} {\bibfield  {journal} {\bibinfo
  {journal} {Physical Review Letters}\ }\textbf {\bibinfo {volume} {110}},\
  \bibinfo {pages} {135702} (\bibinfo {year} {2013})}\BibitemShut {NoStop}\bibitem [{\citenamefont {Giamarchi}\ and\ \citenamefont
  {Tsvelik}(1999)}]{giamarchi_coupled_1999}  \BibitemOpen
  \bibfield  {author} {\bibinfo {author} {\bibfnamefont {T.}~\bibnamefont
  {Giamarchi}}\ and\ \bibinfo {author} {\bibfnamefont {A.~M.}\ \bibnamefont
  {Tsvelik}},\ }\href {\doibase 10.1103/PhysRevB.59.11398} {\bibfield
  {journal} {\bibinfo  {journal} {Physical Review B}\ }\textbf {\bibinfo
  {volume} {59}},\ \bibinfo {pages} {11398} (\bibinfo {year}
  {1999})}\BibitemShut {NoStop}\bibitem [{\citenamefont {Sandvik}(2010)}]{sandvik_computational_2010}  \BibitemOpen
  \bibfield  {author} {\bibinfo {author} {\bibfnamefont {A.~W.}\ \bibnamefont
  {Sandvik}},\ }\href {\doibase doi:10.1063/1.3518900} {\bibfield  {journal}
  {\bibinfo  {journal} {{AIP} Conference Proceedings}\ }\textbf {\bibinfo
  {volume} {1297}},\ \bibinfo {pages} {135} (\bibinfo {year}
  {2010})}\BibitemShut {NoStop}\bibitem [{Note3()}]{Note3}  \BibitemOpen
  \bibinfo {note} {The method is distinct from the Monte Carlo method used in
  studies of entanglement entropies~\cite
  {hastings_measuring_2010,humeniuk_quantum_2012}: there is no ``swap'' of
  configurations involved here, as all replicas are sampled independently
  according to the same density matrix.}\BibitemShut {Stop}\bibitem [{sup()}]{supplementary}  \BibitemOpen
  \href@noop {} {}\bibinfo {note} {Supplementary material}\BibitemShut
  {NoStop}\bibitem [{\citenamefont {Giamarchi}(2004)}]{giamarchi_quantum_2004}  \BibitemOpen
  \bibfield  {author} {\bibinfo {author} {\bibfnamefont {T.}~\bibnamefont
  {Giamarchi}},\ }\href@noop {} {\emph {\bibinfo {title} {Quantum physics in
  one dimension}}}\ (\bibinfo  {publisher} {Clarendon; Oxford University
  Press},\ \bibinfo {address} {Oxford; New York},\ \bibinfo {year}
  {2004})\BibitemShut {NoStop}\bibitem [{\citenamefont {Zaletel}\ \emph {et~al.}(2011)\citenamefont
  {Zaletel}, \citenamefont {Bardarson},\ and\ \citenamefont
  {Moore}}]{zaletel_logarithmic_2011}  \BibitemOpen
  \bibfield  {author} {\bibinfo {author} {\bibfnamefont {M.~P.}\ \bibnamefont
  {Zaletel}}, \bibinfo {author} {\bibfnamefont {J.~H.}\ \bibnamefont
  {Bardarson}}, \ and\ \bibinfo {author} {\bibfnamefont {J.~E.}\ \bibnamefont
  {Moore}},\ }\href {\doibase 10.1103/PhysRevLett.107.020402} {\bibfield
  {journal} {\bibinfo  {journal} {Physical Review Letters}\ }\textbf {\bibinfo
  {volume} {107}},\ \bibinfo {pages} {020402} (\bibinfo {year}
  {2011})}\BibitemShut {NoStop}\bibitem [{\citenamefont {Dagotto}\ and\ \citenamefont
  {Rice}(1996)}]{dagotto_surprises_1996}  \BibitemOpen
  \bibfield  {author} {\bibinfo {author} {\bibfnamefont {E.}~\bibnamefont
  {Dagotto}}\ and\ \bibinfo {author} {\bibfnamefont {T.~M.}\ \bibnamefont
  {Rice}},\ }\href {\doibase 10.1126/science.271.5249.618} {\bibfield
  {journal} {\bibinfo  {journal} {Science}\ }\textbf {\bibinfo {volume}
  {271}},\ \bibinfo {pages} {618} (\bibinfo {year} {1996})}\BibitemShut
  {NoStop}\bibitem [{\citenamefont {R{\"u}egg}\ \emph {et~al.}(2008)\citenamefont
  {R{\"u}egg}, \citenamefont {Kiefer}, \citenamefont {Thielemann},
  \citenamefont {{McMorrow}}, \citenamefont {Zapf}, \citenamefont {Normand},
  \citenamefont {Zvonarev}, \citenamefont {Bouillot}, \citenamefont {Kollath},
  \citenamefont {Giamarchi}, \citenamefont {Capponi}, \citenamefont
  {Poilblanc}, \citenamefont {Biner},\ and\ \citenamefont
  {Kr{\"a}mer}}]{ruegg_thermodynamics_2008}  \BibitemOpen
  \bibfield  {author} {\bibinfo {author} {\bibfnamefont {C.}~\bibnamefont
  {R{\"u}egg}}, \bibinfo {author} {\bibfnamefont {K.}~\bibnamefont {Kiefer}},
  \bibinfo {author} {\bibfnamefont {B.}~\bibnamefont {Thielemann}}, \bibinfo
  {author} {\bibfnamefont {D.~F.}\ \bibnamefont {{McMorrow}}}, \bibinfo
  {author} {\bibfnamefont {V.}~\bibnamefont {Zapf}}, \bibinfo {author}
  {\bibfnamefont {B.}~\bibnamefont {Normand}}, \bibinfo {author} {\bibfnamefont
  {M.~B.}\ \bibnamefont {Zvonarev}}, \bibinfo {author} {\bibfnamefont
  {P.}~\bibnamefont {Bouillot}}, \bibinfo {author} {\bibfnamefont
  {C.}~\bibnamefont {Kollath}}, \bibinfo {author} {\bibfnamefont
  {T.}~\bibnamefont {Giamarchi}}, \bibinfo {author} {\bibfnamefont
  {S.}~\bibnamefont {Capponi}}, \bibinfo {author} {\bibfnamefont
  {D.}~\bibnamefont {Poilblanc}}, \bibinfo {author} {\bibfnamefont
  {D.}~\bibnamefont {Biner}}, \ and\ \bibinfo {author} {\bibfnamefont {K.~W.}\
  \bibnamefont {Kr{\"a}mer}},\ }\href {\doibase 10.1103/PhysRevLett.101.247202}
  {\bibfield  {journal} {\bibinfo  {journal} {Physical Review Letters}\
  }\textbf {\bibinfo {volume} {101}},\ \bibinfo {pages} {247202} (\bibinfo
  {year} {2008})}\BibitemShut {NoStop}\bibitem [{\citenamefont {Klanj{\v s}ek}\ \emph {et~al.}(2008)\citenamefont
  {Klanj{\v s}ek}, \citenamefont {Mayaffre}, \citenamefont {Berthier},
  \citenamefont {Horvati{\'c}}, \citenamefont {Chiari}, \citenamefont
  {Piovesana}, \citenamefont {Bouillot}, \citenamefont {Kollath}, \citenamefont
  {Orignac}, \citenamefont {Citro},\ and\ \citenamefont
  {Giamarchi}}]{klanjsek_controlling_2008}  \BibitemOpen
  \bibfield  {author} {\bibinfo {author} {\bibfnamefont {M.}~\bibnamefont
  {Klanj{\v s}ek}}, \bibinfo {author} {\bibfnamefont {H.}~\bibnamefont
  {Mayaffre}}, \bibinfo {author} {\bibfnamefont {C.}~\bibnamefont {Berthier}},
  \bibinfo {author} {\bibfnamefont {M.}~\bibnamefont {Horvati{\'c}}}, \bibinfo
  {author} {\bibfnamefont {B.}~\bibnamefont {Chiari}}, \bibinfo {author}
  {\bibfnamefont {O.}~\bibnamefont {Piovesana}}, \bibinfo {author}
  {\bibfnamefont {P.}~\bibnamefont {Bouillot}}, \bibinfo {author}
  {\bibfnamefont {C.}~\bibnamefont {Kollath}}, \bibinfo {author} {\bibfnamefont
  {E.}~\bibnamefont {Orignac}}, \bibinfo {author} {\bibfnamefont
  {R.}~\bibnamefont {Citro}}, \ and\ \bibinfo {author} {\bibfnamefont
  {T.}~\bibnamefont {Giamarchi}},\ }\href {\doibase
  10.1103/PhysRevLett.101.137207} {\bibfield  {journal} {\bibinfo  {journal}
  {Physical Review Letters}\ }\textbf {\bibinfo {volume} {101}},\ \bibinfo
  {pages} {137207} (\bibinfo {year} {2008})}\BibitemShut {NoStop}\bibitem [{\citenamefont {Hong}\ \emph {et~al.}(2010)\citenamefont {Hong},
  \citenamefont {Kim}, \citenamefont {Hotta}, \citenamefont {Takano},
  \citenamefont {Tremelling}, \citenamefont {Turnbull}, \citenamefont {Landee},
  \citenamefont {Kang}, \citenamefont {Christensen}, \citenamefont {Lefmann},
  \citenamefont {Schmidt}, \citenamefont {Uhrig},\ and\ \citenamefont
  {Broholm}}]{hong_field-induced_2010}  \BibitemOpen
  \bibfield  {author} {\bibinfo {author} {\bibfnamefont {T.}~\bibnamefont
  {Hong}}, \bibinfo {author} {\bibfnamefont {Y.~H.}\ \bibnamefont {Kim}},
  \bibinfo {author} {\bibfnamefont {C.}~\bibnamefont {Hotta}}, \bibinfo
  {author} {\bibfnamefont {Y.}~\bibnamefont {Takano}}, \bibinfo {author}
  {\bibfnamefont {G.}~\bibnamefont {Tremelling}}, \bibinfo {author}
  {\bibfnamefont {M.~M.}\ \bibnamefont {Turnbull}}, \bibinfo {author}
  {\bibfnamefont {C.~P.}\ \bibnamefont {Landee}}, \bibinfo {author}
  {\bibfnamefont {H.-J.}\ \bibnamefont {Kang}}, \bibinfo {author}
  {\bibfnamefont {N.~B.}\ \bibnamefont {Christensen}}, \bibinfo {author}
  {\bibfnamefont {K.}~\bibnamefont {Lefmann}}, \bibinfo {author} {\bibfnamefont
  {K.~P.}\ \bibnamefont {Schmidt}}, \bibinfo {author} {\bibfnamefont {G.~S.}\
  \bibnamefont {Uhrig}}, \ and\ \bibinfo {author} {\bibfnamefont
  {C.}~\bibnamefont {Broholm}},\ }\href {\doibase
  10.1103/PhysRevLett.105.137207} {\bibfield  {journal} {\bibinfo  {journal}
  {Physical Review Letters}\ }\textbf {\bibinfo {volume} {105}},\ \bibinfo
  {pages} {137207} (\bibinfo {year} {2010})}\BibitemShut {NoStop}\bibitem [{\citenamefont {Jeong}\ \emph {et~al.}(2013)\citenamefont {Jeong},
  \citenamefont {Mayaffre}, \citenamefont {Berthier}, \citenamefont
  {Schmidiger}, \citenamefont {Zheludev},\ and\ \citenamefont
  {Horvati{\'c}}}]{jeong_attractive_2013}  \BibitemOpen
  \bibfield  {author} {\bibinfo {author} {\bibfnamefont {M.}~\bibnamefont
  {Jeong}}, \bibinfo {author} {\bibfnamefont {H.}~\bibnamefont {Mayaffre}},
  \bibinfo {author} {\bibfnamefont {C.}~\bibnamefont {Berthier}}, \bibinfo
  {author} {\bibfnamefont {D.}~\bibnamefont {Schmidiger}}, \bibinfo {author}
  {\bibfnamefont {A.}~\bibnamefont {Zheludev}}, \ and\ \bibinfo {author}
  {\bibfnamefont {M.}~\bibnamefont {Horvati{\'c}}},\ }\href {\doibase
  10.1103/PhysRevLett.111.106404} {\bibfield  {journal} {\bibinfo  {journal}
  {Physical Review Letters}\ }\textbf {\bibinfo {volume} {111}},\ \bibinfo
  {pages} {106404} (\bibinfo {year} {2013})}\BibitemShut {NoStop}\bibitem [{\citenamefont {Bl\"ote}\ and\ \citenamefont {Deng}(2002)}]{Blote}  \BibitemOpen
  \bibfield  {author} {\bibinfo {author} {\bibfnamefont {H.~W.~J.}\
  \bibnamefont {Bl\"ote}}\ and\ \bibinfo {author} {\bibfnamefont
  {Y.}~\bibnamefont {Deng}},\ }\href {\doibase 10.1103/PhysRevE.66.066110}
  {\bibfield  {journal} {\bibinfo  {journal} {Physical Review E}\ }\textbf
  {\bibinfo {volume} {66}},\ \bibinfo {pages} {066110} (\bibinfo {year}
  {2002})}\BibitemShut {NoStop}\bibitem [{\citenamefont {Basko}\ \emph {et~al.}(2006)\citenamefont {Basko},
  \citenamefont {Aleiner},\ and\ \citenamefont {Altshuler}}]{MBL}  \BibitemOpen
  \bibfield  {author} {\bibinfo {author} {\bibfnamefont {D.}~\bibnamefont
  {Basko}}, \bibinfo {author} {\bibfnamefont {I.}~\bibnamefont {Aleiner}}, \
  and\ \bibinfo {author} {\bibfnamefont {B.}~\bibnamefont {Altshuler}},\ }\href
  {\doibase http://dx.doi.org/10.1016/j.aop.2005.11.014} {\bibfield  {journal}
  {\bibinfo  {journal} {Annals of Physics}\ }\textbf {\bibinfo {volume}
  {321}},\ \bibinfo {pages} {1126 } (\bibinfo {year} {2006})}\BibitemShut
  {NoStop}\bibitem [{\citenamefont {Moessner}\ and\ \citenamefont
  {Sondhi}(2001)}]{moessner_resonating_2001}  \BibitemOpen
  \bibfield  {author} {\bibinfo {author} {\bibfnamefont {R.}~\bibnamefont
  {Moessner}}\ and\ \bibinfo {author} {\bibfnamefont {S.~L.}\ \bibnamefont
  {Sondhi}},\ }\href {\doibase 10.1103/PhysRevLett.86.1881} {\bibfield
  {journal} {\bibinfo  {journal} {Physical Review Letters}\ }\textbf {\bibinfo
  {volume} {86}},\ \bibinfo {pages} {1881} (\bibinfo {year}
  {2001})}\BibitemShut {NoStop}\bibitem [{\citenamefont {Misguich}\ \emph {et~al.}(2002)\citenamefont
  {Misguich}, \citenamefont {Serban},\ and\ \citenamefont
  {Pasquier}}]{misguich_quantum_2002}  \BibitemOpen
  \bibfield  {author} {\bibinfo {author} {\bibfnamefont {G.}~\bibnamefont
  {Misguich}}, \bibinfo {author} {\bibfnamefont {D.}~\bibnamefont {Serban}}, \
  and\ \bibinfo {author} {\bibfnamefont {V.}~\bibnamefont {Pasquier}},\ }\href
  {\doibase 10.1103/PhysRevLett.89.137202} {\bibfield  {journal} {\bibinfo
  {journal} {Physical Review Letters}\ }\textbf {\bibinfo {volume} {89}},\
  \bibinfo {pages} {137202} (\bibinfo {year} {2002})}\BibitemShut {NoStop}\bibitem [{\citenamefont {Levin}\ and\ \citenamefont
  {Wen}(2005)}]{levin_string-net_2005}  \BibitemOpen
  \bibfield  {author} {\bibinfo {author} {\bibfnamefont {M.~A.}\ \bibnamefont
  {Levin}}\ and\ \bibinfo {author} {\bibfnamefont {X.-G.}\ \bibnamefont
  {Wen}},\ }\href {\doibase 10.1103/PhysRevB.71.045110} {\bibfield  {journal}
  {\bibinfo  {journal} {Physical Review B}\ }\textbf {\bibinfo {volume} {71}},\
  \bibinfo {pages} {045110} (\bibinfo {year} {2005})}\BibitemShut {NoStop}\bibitem [{\citenamefont {Hermanns}\ and\ \citenamefont
  {Trebst}(2013)}]{hermanns_renyi_2013}  \BibitemOpen
  \bibfield  {author} {\bibinfo {author} {\bibfnamefont {M.}~\bibnamefont
  {Hermanns}}\ and\ \bibinfo {author} {\bibfnamefont {S.}~\bibnamefont
  {Trebst}},\ }\href {http://arxiv.org/abs/1309.3793} {\bibfield  {journal}
  {\bibinfo  {journal} {{arXiv:1309.3793} [cond-mat]}\ } (\bibinfo {year}
  {2013})}\BibitemShut {NoStop}\bibitem [{\citenamefont {Castelnovo}\ and\ \citenamefont
  {Chamon}(2007)}]{castelnovo_topological_2007}  \BibitemOpen
  \bibfield  {author} {\bibinfo {author} {\bibfnamefont {C.}~\bibnamefont
  {Castelnovo}}\ and\ \bibinfo {author} {\bibfnamefont {C.}~\bibnamefont
  {Chamon}},\ }\href {\doibase 10.1103/PhysRevB.76.174416} {\bibfield
  {journal} {\bibinfo  {journal} {Physical Review B}\ }\textbf {\bibinfo
  {volume} {76}},\ \bibinfo {pages} {174416} (\bibinfo {year}
  {2007})}\BibitemShut {NoStop}\bibitem [{\citenamefont {Albuquerque}\ \emph {et~al.}(2007)\citenamefont
  {Albuquerque}, \citenamefont {Alet}, \citenamefont {Corboz}, \citenamefont
  {Dayal}, \citenamefont {Feiguin}, \citenamefont {Fuchs}, \citenamefont
  {Gamper}, \citenamefont {Gull}, \citenamefont {Gürtler}, \citenamefont
  {Honecker}, \citenamefont {Igarashi}, \citenamefont {Körner}, \citenamefont
  {Kozhevnikov}, \citenamefont {L{\"a}uchli}, \citenamefont {Manmana},
  \citenamefont {Matsumoto}, \citenamefont {McCulloch}, \citenamefont {Michel},
  \citenamefont {Noack}, \citenamefont {Pawłowski}, \citenamefont {Pollet},
  \citenamefont {Pruschke}, \citenamefont {Schollw{\"o}ck}, \citenamefont
  {Todo}, \citenamefont {Trebst}, \citenamefont {Troyer}, \citenamefont
  {Werner},\ and\ \citenamefont {Wessel}}]{ALPS13}  \BibitemOpen
  \bibfield  {author} {\bibinfo {author} {\bibfnamefont {A.}~\bibnamefont
  {Albuquerque}}, \bibinfo {author} {\bibfnamefont {F.}~\bibnamefont {Alet}},
  \bibinfo {author} {\bibfnamefont {P.}~\bibnamefont {Corboz}}, \bibinfo
  {author} {\bibfnamefont {P.}~\bibnamefont {Dayal}}, \bibinfo {author}
  {\bibfnamefont {A.}~\bibnamefont {Feiguin}}, \bibinfo {author} {\bibfnamefont
  {S.}~\bibnamefont {Fuchs}}, \bibinfo {author} {\bibfnamefont
  {L.}~\bibnamefont {Gamper}}, \bibinfo {author} {\bibfnamefont
  {E.}~\bibnamefont {Gull}}, \bibinfo {author} {\bibfnamefont {S.}~\bibnamefont
  {Gürtler}}, \bibinfo {author} {\bibfnamefont {A.}~\bibnamefont {Honecker}},
  \bibinfo {author} {\bibfnamefont {R.}~\bibnamefont {Igarashi}}, \bibinfo
  {author} {\bibfnamefont {M.}~\bibnamefont {Körner}}, \bibinfo {author}
  {\bibfnamefont {A.}~\bibnamefont {Kozhevnikov}}, \bibinfo {author}
  {\bibfnamefont {A.}~\bibnamefont {L{\"a}uchli}}, \bibinfo {author}
  {\bibfnamefont {S.}~\bibnamefont {Manmana}}, \bibinfo {author} {\bibfnamefont
  {M.}~\bibnamefont {Matsumoto}}, \bibinfo {author} {\bibfnamefont
  {I.}~\bibnamefont {McCulloch}}, \bibinfo {author} {\bibfnamefont
  {F.}~\bibnamefont {Michel}}, \bibinfo {author} {\bibfnamefont
  {R.}~\bibnamefont {Noack}}, \bibinfo {author} {\bibfnamefont
  {G.}~\bibnamefont {Pawłowski}}, \bibinfo {author} {\bibfnamefont
  {L.}~\bibnamefont {Pollet}}, \bibinfo {author} {\bibfnamefont
  {T.}~\bibnamefont {Pruschke}}, \bibinfo {author} {\bibfnamefont
  {U.}~\bibnamefont {Schollw{\"o}ck}}, \bibinfo {author} {\bibfnamefont
  {S.}~\bibnamefont {Todo}}, \bibinfo {author} {\bibfnamefont {S.}~\bibnamefont
  {Trebst}}, \bibinfo {author} {\bibfnamefont {M.}~\bibnamefont {Troyer}},
  \bibinfo {author} {\bibfnamefont {P.}~\bibnamefont {Werner}}, \ and\ \bibinfo
  {author} {\bibfnamefont {S.}~\bibnamefont {Wessel}},\ }\href {\doibase
  http://dx.doi.org/10.1016/j.jmmm.2006.10.304} {\bibfield  {journal} {\bibinfo
   {journal} {Journal of Magnetism and Magnetic Materials}\ }\textbf {\bibinfo
  {volume} {310}},\ \bibinfo {pages} {1187 } (\bibinfo {year}
  {2007})}\BibitemShut {NoStop}\bibitem [{\citenamefont {Bauer}\ \emph {et~al.}(2011)\citenamefont {Bauer},
  \citenamefont {Carr}, \citenamefont {Evertz}, \citenamefont {Feiguin},
  \citenamefont {Freire}, \citenamefont {Fuchs}, \citenamefont {Gamper},
  \citenamefont {Gukelberger}, \citenamefont {Gull}, \citenamefont {Guertler},
  \citenamefont {Hehn}, \citenamefont {Igarashi}, \citenamefont {Isakov},
  \citenamefont {Koop}, \citenamefont {Ma}, \citenamefont {Mates},
  \citenamefont {Matsuo}, \citenamefont {Parcollet}, \citenamefont
  {Pawłowski}, \citenamefont {Picon}, \citenamefont {Pollet}, \citenamefont
  {Santos}, \citenamefont {Scarola}, \citenamefont {Schollw{\"o}ck},
  \citenamefont {Silva}, \citenamefont {Surer}, \citenamefont {Todo},
  \citenamefont {Trebst}, \citenamefont {Troyer}, \citenamefont {Wall},
  \citenamefont {Werner},\ and\ \citenamefont {Wessel}}]{ALPS2}  \BibitemOpen
  \bibfield  {author} {\bibinfo {author} {\bibfnamefont {B.}~\bibnamefont
  {Bauer}}, \bibinfo {author} {\bibfnamefont {L.~D.}\ \bibnamefont {Carr}},
  \bibinfo {author} {\bibfnamefont {H.~G.}\ \bibnamefont {Evertz}}, \bibinfo
  {author} {\bibfnamefont {A.}~\bibnamefont {Feiguin}}, \bibinfo {author}
  {\bibfnamefont {J.}~\bibnamefont {Freire}}, \bibinfo {author} {\bibfnamefont
  {S.}~\bibnamefont {Fuchs}}, \bibinfo {author} {\bibfnamefont
  {L.}~\bibnamefont {Gamper}}, \bibinfo {author} {\bibfnamefont
  {J.}~\bibnamefont {Gukelberger}}, \bibinfo {author} {\bibfnamefont
  {E.}~\bibnamefont {Gull}}, \bibinfo {author} {\bibfnamefont {S.}~\bibnamefont
  {Guertler}}, \bibinfo {author} {\bibfnamefont {A.}~\bibnamefont {Hehn}},
  \bibinfo {author} {\bibfnamefont {R.}~\bibnamefont {Igarashi}}, \bibinfo
  {author} {\bibfnamefont {S.~V.}\ \bibnamefont {Isakov}}, \bibinfo {author}
  {\bibfnamefont {D.}~\bibnamefont {Koop}}, \bibinfo {author} {\bibfnamefont
  {P.~N.}\ \bibnamefont {Ma}}, \bibinfo {author} {\bibfnamefont
  {P.}~\bibnamefont {Mates}}, \bibinfo {author} {\bibfnamefont
  {H.}~\bibnamefont {Matsuo}}, \bibinfo {author} {\bibfnamefont
  {O.}~\bibnamefont {Parcollet}}, \bibinfo {author} {\bibfnamefont
  {G.}~\bibnamefont {Pawłowski}}, \bibinfo {author} {\bibfnamefont {J.~D.}\
  \bibnamefont {Picon}}, \bibinfo {author} {\bibfnamefont {L.}~\bibnamefont
  {Pollet}}, \bibinfo {author} {\bibfnamefont {E.}~\bibnamefont {Santos}},
  \bibinfo {author} {\bibfnamefont {V.~W.}\ \bibnamefont {Scarola}}, \bibinfo
  {author} {\bibfnamefont {U.}~\bibnamefont {Schollw{\"o}ck}}, \bibinfo
  {author} {\bibfnamefont {C.}~\bibnamefont {Silva}}, \bibinfo {author}
  {\bibfnamefont {B.}~\bibnamefont {Surer}}, \bibinfo {author} {\bibfnamefont
  {S.}~\bibnamefont {Todo}}, \bibinfo {author} {\bibfnamefont {S.}~\bibnamefont
  {Trebst}}, \bibinfo {author} {\bibfnamefont {M.}~\bibnamefont {Troyer}},
  \bibinfo {author} {\bibfnamefont {M.~L.}\ \bibnamefont {Wall}}, \bibinfo
  {author} {\bibfnamefont {P.}~\bibnamefont {Werner}}, \ and\ \bibinfo {author}
  {\bibfnamefont {S.}~\bibnamefont {Wessel}},\ }\href
  {http://stacks.iop.org/1742-5468/2011/i=05/a=P05001} {\bibfield  {journal}
  {\bibinfo  {journal} {Journal of Statistical Mechanics: Theory and
  Experiment}\ }\textbf {\bibinfo {volume} {2011}},\ \bibinfo {pages} {P05001}
  (\bibinfo {year} {2011})}\BibitemShut {NoStop}\bibitem [{\citenamefont {Hastings}\ \emph {et~al.}(2010)\citenamefont
  {Hastings}, \citenamefont {Gonz{\'a}lez}, \citenamefont {Kallin},\ and\
  \citenamefont {Melko}}]{hastings_measuring_2010}  \BibitemOpen
  \bibfield  {author} {\bibinfo {author} {\bibfnamefont {M.~B.}\ \bibnamefont
  {Hastings}}, \bibinfo {author} {\bibfnamefont {I.}~\bibnamefont
  {Gonz{\'a}lez}}, \bibinfo {author} {\bibfnamefont {A.~B.}\ \bibnamefont
  {Kallin}}, \ and\ \bibinfo {author} {\bibfnamefont {R.~G.}\ \bibnamefont
  {Melko}},\ }\href {\doibase 10.1103/PhysRevLett.104.157201} {\bibfield
  {journal} {\bibinfo  {journal} {Physical Review Letters}\ }\textbf {\bibinfo
  {volume} {104}},\ \bibinfo {pages} {157201} (\bibinfo {year}
  {2010})}\BibitemShut {NoStop}\bibitem [{\citenamefont {Humeniuk}\ and\ \citenamefont
  {Roscilde}(2012)}]{humeniuk_quantum_2012}  \BibitemOpen
  \bibfield  {author} {\bibinfo {author} {\bibfnamefont {S.}~\bibnamefont
  {Humeniuk}}\ and\ \bibinfo {author} {\bibfnamefont {T.}~\bibnamefont
  {Roscilde}},\ }\href {\doibase 10.1103/PhysRevB.86.235116} {\bibfield
  {journal} {\bibinfo  {journal} {Physical Review B}\ }\textbf {\bibinfo
  {volume} {86}},\ \bibinfo {pages} {235116} (\bibinfo {year}
  {2012})}\BibitemShut {NoStop}\end{thebibliography}

\begin{thebibliography}{10}
\expandafter\ifx\csname url\endcsname\relax
  \def\url#1{\texttt{#1}}\fi
\expandafter\ifx\csname urlprefix\endcsname\relax\def\urlprefix{URL }\fi
\providecommand{\bibinfo}[2]{#2}
\providecommand{\eprint}[2][]{\url{#2}}

\bibitem{stephan_renyi_2010}
\bibinfo{author}{St{\'e}phan, J.-M.}, \bibinfo{author}{Misguich, G.} \&
  \bibinfo{author}{Pasquier, V.}
\newblock \bibinfo{title}{R{\'e}nyi entropy of a line in two-dimensional ising
  models}.
\newblock \emph{\bibinfo{journal}{Physical Review B}}
  \textbf{\bibinfo{volume}{82}}, \bibinfo{pages}{125455}
  (\bibinfo{year}{2010}).
\newblock \urlprefix\url{http://link.aps.org/doi/10.1103/PhysRevB.82.125455}.

\bibitem{Blote}
\bibinfo{author}{Bl\"ote, H. W.~J.} \& \bibinfo{author}{Deng, Y.}
\newblock \bibinfo{title}{Cluster monte carlo simulation of the transverse
  ising model}.
\newblock \emph{\bibinfo{journal}{Physical Review E}}
  \textbf{\bibinfo{volume}{66}}, \bibinfo{pages}{066110}
  (\bibinfo{year}{2002}).
\newblock \urlprefix\url{http://link.aps.org/doi/10.1103/PhysRevE.66.066110}.

\bibitem{young_everything_2012}
\bibinfo{author}{Young, P.}
\newblock \bibinfo{title}{Everything you wanted to know about data analysis and
  fitting but were afraid to ask}.
\newblock \bibinfo{type}{{arXiv} e-print} \bibinfo{number}{1210.3781}
  (\bibinfo{year}{2012}).
\newblock \urlprefix\url{http://arxiv.org/abs/1210.3781}.

\bibitem{lieb_ordering_1962}
\bibinfo{author}{Lieb, E.} \& \bibinfo{author}{Mattis, D.}
\newblock \bibinfo{title}{Ordering energy levels of interacting spin systems}.
\newblock \emph{\bibinfo{journal}{Journal of Mathematical Physics}}
  \textbf{\bibinfo{volume}{3}}, \bibinfo{pages}{749--751}
  (\bibinfo{year}{1962}).
\newblock \urlprefix\url{http://jmp.aip.org/resource/1/jmapaq/v3/i4/p749_s1}.

\bibitem{Anderson}
\bibinfo{author}{Anderson, P.~W.}
\newblock \bibinfo{title}{An approximate quantum theory of the
  antiferromagnetic ground state}.
\newblock \emph{\bibinfo{journal}{Physical Review}}
  \textbf{\bibinfo{volume}{86}}, \bibinfo{pages}{694--701}
  (\bibinfo{year}{1952}).
\newblock \urlprefix\url{http://link.aps.org/doi/10.1103/PhysRev.86.694}.

\bibitem{lhuillier_frustrated_2005}
\bibinfo{author}{Lhuillier, C.}
\newblock \bibinfo{title}{Frustrated quantum magnets}.
\newblock \bibinfo{type}{{arXiv} e-print} \bibinfo{number}{cond-mat/0502464}
  (\bibinfo{year}{2005}).
\newblock \urlprefix\url{http://arxiv.org/abs/cond-mat/0502464}.

\bibitem{google_sparsehash}
\bibinfo{howpublished}{\texttt{http://code.google.com/p/sparsehash/}}.

\bibitem{Sandvik03}
\bibinfo{author}{Sandvik, A.~W.}
\newblock \bibinfo{title}{Stochastic series expansion method for quantum ising
  models with arbitrary interactions}.
\newblock \emph{\bibinfo{journal}{Physical Review E}}
  \textbf{\bibinfo{volume}{68}}, \bibinfo{pages}{056701}
  (\bibinfo{year}{2003}).
\newblock \urlprefix\url{http://link.aps.org/doi/10.1103/PhysRevE.68.056701}.

\bibitem{Albuquerque10}
\bibinfo{author}{Albuquerque, A.~F.}, \bibinfo{author}{Alet, F.},
  \bibinfo{author}{Sire, C.} \& \bibinfo{author}{Capponi, S.}
\newblock \bibinfo{title}{Quantum critical scaling of fidelity susceptibility}.
\newblock \emph{\bibinfo{journal}{Physical Review B}}
  \textbf{\bibinfo{volume}{81}}, \bibinfo{pages}{064418}
  (\bibinfo{year}{2010}).
\newblock \urlprefix\url{http://link.aps.org/doi/10.1103/PhysRevB.81.064418}.

\bibitem{Sandvik99}
\bibinfo{author}{Sandvik, A.~W.} \& \bibinfo{author}{Hamer, C.~J.}
\newblock \bibinfo{title}{Ground-state parameters, finite-size scaling, and
  low-temperature properties of the two-dimensional $s=\frac{1}{2}$
  $\mathrm{XY}$ model}.
\newblock \emph{\bibinfo{journal}{Physical Review B}}
  \textbf{\bibinfo{volume}{60}}, \bibinfo{pages}{6588--6593}
  (\bibinfo{year}{1999}).
\newblock \urlprefix\url{http://link.aps.org/doi/10.1103/PhysRevB.60.6588}.

\bibitem{sandvik_computational_2010}
\bibinfo{author}{Sandvik, A.~W.}
\newblock \bibinfo{title}{Computational studies of quantum spin systems}.
\newblock \emph{\bibinfo{journal}{{AIP} Conference Proceedings}}
  \textbf{\bibinfo{volume}{1297}}, \bibinfo{pages}{135--338}
  (\bibinfo{year}{2010}).
\newblock
  \urlprefix\url{http://proceedings.aip.org/resource/2/apcpcs/1297/1/135_1}.

\end{thebibliography}
\end{document}